\documentclass{aa}

\usepackage{graphicx}
\usepackage[varg]{txfonts}
\usepackage{longtable}
\usepackage{lscape}
\usepackage{natbib}

\newcommand{\feh}{\mathrm{[Fe/H]}}
\newcommand{\teff}{T_\mathrm{eff}}
\newcommand{\logg}{\log g}
\newcommand{\vt}{v_t}

\newcommand{\fei}{Fe\,\textsc{i}}
\newcommand{\feii}{Fe\,\textsc{ii}}
\newcommand{\kms}{km\,s$^{-1}$}
\newcommand{\ha}{\mathrm{H}\alpha}

\begin{document} 

\title{The Solar Twin Planet Search}
\subtitle{I. Fundamental parameters of the stellar sample}

\titlerunning{Fundamental parameters of solar twins}

\author{I. Ram\'irez\inst{1}   \and
        J. Mel\'endez\inst{2}  \and
        J. Bean\inst{3}        \and
        M. Asplund\inst{4}     \and
        M. Bedell\inst{3}      \and
        T. Monroe\inst{2}      \and
        L. Casagrande\inst{4}  \and \\
        L. Schirbel\inst{2}    \and
        S. Dreizler\inst{5}    \and
        J. Teske\thanks{Carnegie Origins Fellow}\inst{6,7,8} \and
        M. Tucci Maia\inst{2}  \and
        A. Alves-Brito\inst{9} \and
        P. Baumann\inst{10}
       }

\institute{McDonald Observatory and Department of Astronomy,
           University of Texas at Austin, USA\\
           \email{ivan@astro.as.utexas.edu}
           \and
           Departamento de Astronomia do IAG/USP,
           Universidade de S\~ao Paulo, Brazil
           \and
           Department of Astronomy and Astrophysics,
           University of Chicago, USA
           \and
           Research School of Astronomy and Astrophysics,
           Mount Stromlo Observatory,
           The Australian National University, Australia
           \and
           Institut f\"{u}r Astrophysik,
           University of G\"{o}ttingen, Germany
           \and
           Steward Observatory, Department of Astronomy,
           University of Arizona, USA
           \and
           Department of Terrestrial Magnetism,
           Carnegie Institution of Washington,
           USA
           \and
           The Observatories of the Carnegie Institution For Science,
           Pasadena, California, USA
           \and
           Instituto de Fisica,
           Universidade Federal do Rio Grande do Sul,
           Porto Alegre, RS, Brazil
           \and
           Unaffiliated
          }

\date{Received --- --, ---; accepted --- --, ---}

\abstract
{We are carrying out a search for planets around a sample of solar twin stars using the HARPS spectrograph. The goal of this project is to exploit the advantage offered by solar twins to obtain chemical abundances of unmatched precision. This survey will enable new studies of the stellar composition -- planet connection.}
{We determine the fundamental parameters of the 88 solar twin stars that have been chosen as targets for our experiment.}
{We used the MIKE spectrograph on the Magellan Clay Telescope to acquire high resolution, high signal-to-noise ratio spectra of our sample stars. We measured the equivalent widths of iron lines and used strict differential excitation/ionization balance analysis to determine atmospheric parameters of unprecedented internal precision: $\sigma(\teff)=7$\,K, $\sigma(\logg)=0.019$, $\sigma(\feh)=0.006$\,dex, $\sigma(\vt)=0.016$\,\kms. Reliable relative ages and highly precise masses were then estimated using theoretical isochrones.}
{The spectroscopic parameters we derived are in good agreement with those measured using other independent techniques. There is even better agreement if the sample is restricted to those stars with the most internally precise determinations of stellar parameters in every technique involved. The root-mean-square scatter of the differences seen is fully compatible with the observational errors, demonstrating, as assumed thus far, that systematic uncertainties in the stellar parameters are negligible in the study of solar twins. We find a tight activity--age relation for our sample stars, which validates the internal precision of our dating method. Furthermore, we find that the solar cycle is perfectly consistent both with this trend and its star-to-star scatter.}
{We present the largest sample of solar twins analyzed homogeneously using high quality spectra. The fundamental parameters derived from this work will be employed in subsequent work that aims to explore the connections between planet formation and stellar chemical composition.}


\maketitle

\section{Introduction}

Planets form by sequestering refractory and volatile material from protoplanetary disks. This process may affect the chemical composition of the gas accreted during the final stages of star formation. Therefore, it can potentially imprint its signatures on the composition of the outermost layers of the host stars. Also, the composition of the nebula that stars and their accompanying planetary systems form out of may influence the number and type of resulting planets. Thus, there may be a connection between the chemical composition of stars and the presence and composition of different types of planets.

The classic example of the relationship between stellar abundances and planets is the observed higher frequency of giant planets around stars of higher metallicity \cite[e.g.,][]{gonzalez97,santos04,valenti05,ghezzi10:metallicity}. Other signatures of planet formation are harder to detect because they are expected to be at the 1\,\% level, or lower \citep{chambers10}. Nevertheless, this level of precision can be achieved by studying solar twins \citep{cayrel96}, stars which are spectroscopically very similar to the Sun. This is because the many systematic effects that plague classical elemental abundance determinations can be eliminated or minimized by a strict differential analysis between the solar twins and the Sun.

In the past few years, the study of solar twins has revealed three potential signatures of planet formation in addition to the planet-metallicity correlation:

\begin{itemize}

\item i) a deficiency of about 0.1\,dex\footnote{In the standard elemental abundance scale: $[X/H]=A_X-A_X^\odot$, where $A_X=\log(n_X/n_H)+12$ and $n_X$ is the number density of $X$ nuclei in the stellar photosphere.} in refractory material relative to volatiles in the Sun when compared to solar twins \citep{melendez09:twins,ramirez09,ramirez10}, with a trend with condensation temperature that could be explained by material with Earth and meteoritic composition \citep{chambers10}, hence suggesting a signature of terrestrial planet formation (see also \citealt{gonzalez-hernandez10,gonzalez-hernandez13,gonzalez10,gonzalez11,schuler11,melendez12});

\item ii) a nearly constant offset of about 0.04\,dex in elemental abundances between the solar analog components of the 16\,Cygni binary system \citep{laws01,ramirez11,tucci14}, where the secondary hosts a giant planet but no planet has been detected so far around the primary (see also \citealt{schuler11:16cyg});

\item iii) on top of the roughly constant offset between the abundances of 16\,Cygni\,A and B, there are additional differences (of order 0.015\,dex) for the refractories, with a condensation temperature trend that can be attributed to the rocky accretion core of the giant planet 16\,Cygni\,B\,b \citep{tucci14}.

\end{itemize}

In order to explore the connection between chemical abundance anomalies and planet architecture further, we have an ongoing Large ESO Program (188.C-0265, P.I.~J.\,Mel\'endez) to characterize planets around solar twins using the HARPS spectrograph, the world’s most powerful ground-based planet-hunting machine \cite[e.g.,][]{mayor03}. We will exploit the synergy between the high precision in radial velocities that can be achieved by HARPS ($\sim1$\,m\,s$^{-1}$)\footnote{This value of instrumental precision has in fact been confirmed in our existing HARPS data for the least active star.} and the high precision in chemical abundances that can be obtained in solar twins ($\sim0.01$\,dex). This project is described in more detail in Section~\ref{s:project}.

In this paper we present our sample and determine a homogeneous set of precise fundamental stellar parameters using complementary high resolution, high signal-to-noise ratio spectroscopic observations of solar twins and the Sun. In addition, we verify the results with stellar parameters obtained through other techniques. Also, stellar activity indices, masses, and ages are provided. The fundamental stellar properties here derived will be used in a series of forthcoming papers on the detailed chemical composition of our solar twin sample and on the characterization of their planets with HARPS.

\section{HARPS planet search around solar twins} \label{s:project}

Our HARPS Large Program includes about 60 solar twins, significantly expanding
the hunt for planets around stars closely resembling the
Sun. This project fully exploits the advantage offered by solar twins to
study stellar composition with unprecedented precision in synergy with the
superior planet hunting capabilities offered by HARPS.

In the planet search program we aim for a uniform and deep
characterization of the planetary systems orbiting solar twin stars. This
means not only finding out what planets exist around these stars, but also
what planets do not exist. We have designed our program so that we can
obtain a consistent level of sensitivity for all the stars, to put strict
constraints on the nature of their orbiting planets. We emphasize that we
are not setting out to detect Earth twins (this is about one order of
magnitude beyond the present-day capabilities of HARPS) but we will make as
complete an inventory of the planetary systems as is possible with current
instruments in order to search for correlations between abundance
signature and planet properties.

Our HARPS Large Program started in October 2011 and it will last four
years, with 22 nights per year that are broken up into two runs of seven
nights and two runs of four nights. Our simulations of planet
detectability suggest that a long run per semester aids in finding
low-mass planets, while a second shorter run improves sampling of longer
period planets and helps eliminate blind spots that could arise from
aliasing. We set the minimum exposure times to what is necessary to
achieve a photon-limited precision of 1\,m\,s$^{-1}$ or 15 minutes,
whichever is the longest. The motivation for using 15 minute minimum total
exposure times for a visit is to average the five-minute $p$-mode
oscillations of Sun-like stars to below 1\,m\,s$^{-1}$
\citep[e.g.,][]{mayor03,lovis06,dumusque11}.
For the brightest stars in our sample we take
multiple shorter exposures over 15 minutes to avoid saturating the
detector. Simulations indicate that the precision, sampling, and total
number of measurements from our program will allow us to be sensitive to
planets with masses down to the super-Earth regime (i.e. $<
10\,M_{\oplus}$) in short period orbits (up to 10 days),
the ice giant regime (i.e. 10 -- 25\,$M_{\oplus}$) in
intermediate period orbits (up to 100 days), and the gas giant regime
in long period orbits (100 days or more).

\begin{figure}
\centering
\includegraphics[bb=0 10 240 660,width=8.5cm]{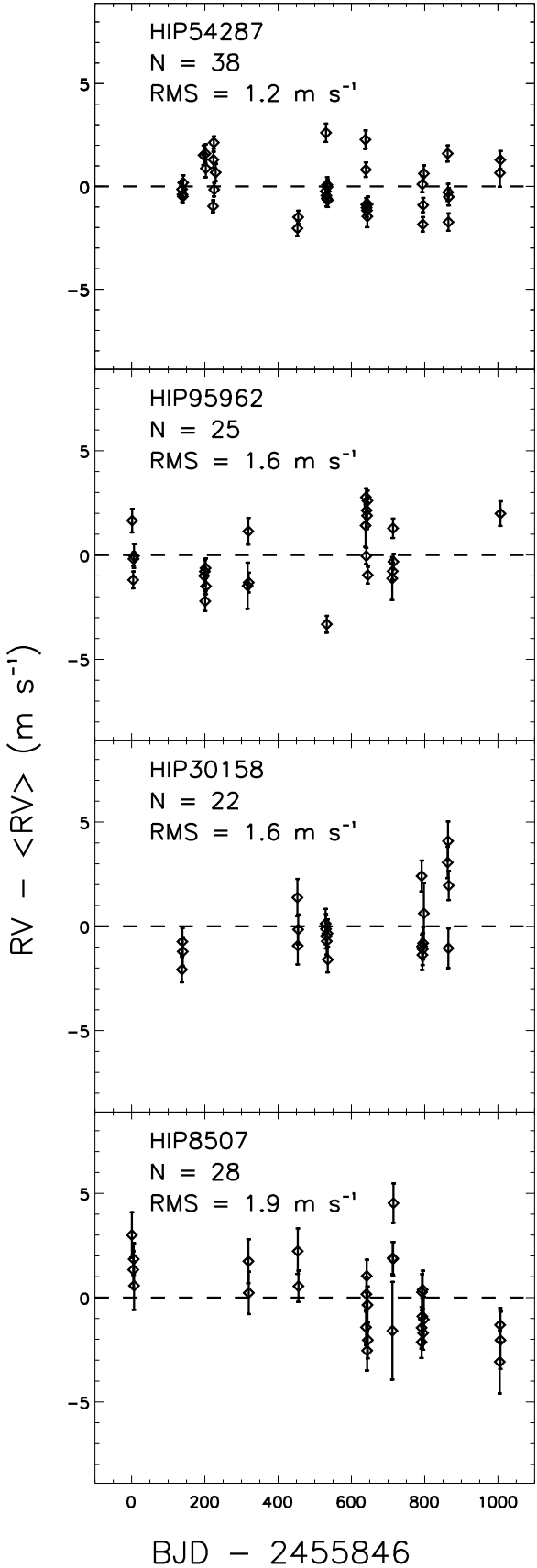}
\caption{Radial velocities measured with HARPS for four of the solar twins in our sample. Dashed lines correspond to each star's average radial velocity (RV) value. The low RMS scatters and lack of apparent instrumental trends in RV demonstrate the high precision achieved with HARPS.}
\label{f:constantrv}
\end{figure}

\begin{figure}
\centering
\includegraphics[bb=0 10 470 376,width=8.5cm]{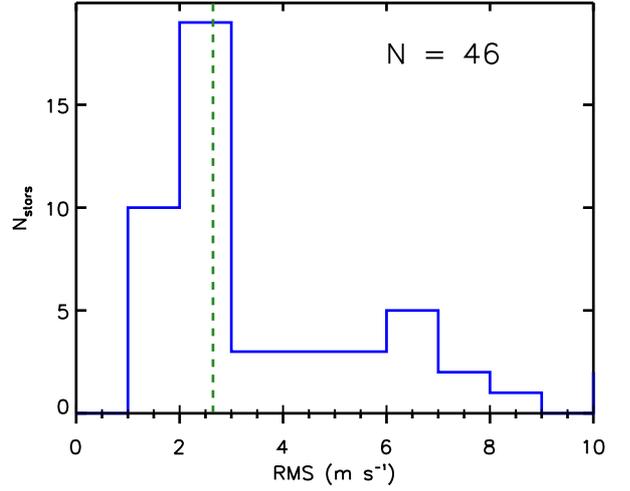}
\caption{Distribution of RMS scatter in radial velocities for 46 of the solar twins currently being monitored with HARPS. An additional 15 stars being monitored have RMS ranging from 10 to 230 m\,s$^{-1}$.  The dashed line corresponds to the median RMS of the 46 stars shown.}
\label{f:rms}
\end{figure}

In Figure~\ref{f:constantrv}, we show the radial velocities of four stars with very low levels of radial velocity variability, corroborating that a precision of 1\,m\,s$^{-1}$ can be achieved. In Figure~\ref{f:rms} we show the RMS (root-mean-squared scatter) radial velocities for all the low variability stars in our sample. Several stars in the sample show clear radial velocity variations. Some of these variations likely correspond to planets and will be the subject of future papers in this series.

\section{Data}

\subsection{Sample selection}

Our sample stars were chosen first from our previous dedicated searches for solar twins at the McDonald \citep{melendez07:twins,ramirez09} and Las Campanas \citep{melendez09:twins} observatories. Those searches were mainly based on measured colors and parallaxes, by matching within the error bars both the solar colors (preliminary values of those given in \citealt{melendez10}, \citealt{ramirez12_suncolor}, and \citealt{casagrande12}) and the Sun’s absolute magnitude. We also added more targets from our spectroscopic analysis of solar twins from the S$^4$N database \citep{allende04:s4n} and the HARPS/ESO archive, as reported in \cite{baumann10}. Finally, we selected additional solar twins from the large samples of \cite{valenti05} and \cite{bensby14}. For the latter two cases we selected stars within 100\,K in $\teff$, 0.1\,dex in $\logg$, and 0.1\,dex in [Fe/H] from the Sun’s values.

We selected a total of 88 solar twins for homogeneous high resolution spectroscopic observations. The sample is presented in Table~\ref{t:sample}. From this sample, about 60 stars are being observed in our HARPS planet search project; the rest have been already characterized by other planet search programs.

\subsection{Spectroscopic observations} \label{s:observations}

The high-quality spectra employed for the stellar parameter and abundance analysis in this paper were acquired with the MIKE spectrograph \citep{bernstein03} on the 6.5\,m Clay Magellan Telescope at Las Campanas Observatory. We observed our targets in five different runs between January 2011 and May 2012 (Table~\ref{t:reference}). The observational setup (described below) was identical in all these runs.

We used the standard MIKE setup, which fully covers the wavelength range from 320 to 1\,000\,nm, and the 0.35\,arcsec (width) slit, which results in a spectral resolution $R=\lambda/\Delta\lambda=83\,000$ (65\,000) in the blue (red) CCD. We targeted a signal-to-noise ratio ($S/N$) per-pixel of at least 400 at 6000\,\AA\ in order to obtain spectra of quality similar to that used by \cite{melendez09:twins}, who were the first to detect the proposed chemical signature of terrestrial planet formation. Multiple consecutive exposures of each of our targets were taken to reach this very high $S/N$ requirement.

The spectra were reduced with the CarnegiePython MIKE pipeline,\footnote{\url{http://code.obs.carnegiescience.edu/mike}} which trims the image and corrects for overscan, applies the flat fields (both lamp and ``milky'')\footnote{Milky flats are blurred flat-field images that illuminate well the gaps between orders and are used to better correct the order edges.} to the object images, removes scattered light and subtracts sky background. It proceeds to extract the stellar flux order-by-order, and it applies a wavelength mapping based on Th-Ar lamp exposures taken every 2-3 hours during each night. Finally, it co-adds multiple exposures of the same target. Hereafter, we refer to these data as the ``extracted'' spectra.

We used IRAF's\footnote{IRAF is distributed by the National Optical Astronomy Observatory, which is operated by the Association of Universities for Research in Astronomy (AURA) under cooperative agreement with the National Science Foundation.} \texttt{dopcor} and \texttt{rvcor} tasks to compute and correct for barycentric motion as well as the stars' absolute radial velocities. The latter were derived as described in Section~\ref{s:rv}. Each spectral order was then continuum normalized using 12th order polynomial fits to the upper envelopes of the data. We excluded $\sim100$ pixels on the edges of each order to avoid their low counts and overall negative impact on the continuum normalization. This did not compromise the continuous wavelength coverage of our data. However, we excluded the 5 reddest orders of the red CCD and discarded the 19 bluest order of the blue CCD. The reason for this is that our continuum normalization is not reliable beyond these limits and we found no need to include those wavelengths for our present purposes. Thus, the wavelength coverage of these spectra is reduced to $\simeq4000-8000\,\AA$. Finally, IRAF's \texttt{scombine} task was used to merge all orders and create a final single-column FITS spectrum for each star. Hereafter, these data are referred to as the ``normalized'' spectra.

\begin{table}
\small\centering
\caption{Observing runs and reference star observations}
\label{t:reference}
\begin{tabular}{clrl}\hline\hline
Run & Dates & Spectra\tablefootmark{1} & Reference(s) \\ \hline
1 & 1--4 Jan 2011       & 35 & Iris \\
2 & 23--24 Jun 2011        & 13 & Vesta and 18\,Sco \\
3 & 9--10 Sep 2011    & 18 & Vesta \\
4 & 23 Feb 2012        &  4 & 18\,Sco \\
5 & 29-30 Apr, 1 May 2012 & 17 & 18\,Sco ($\times2$) \\
\hline
\end{tabular}
\tablefoottext{1}{Number of spectra acquired other than that of the reference target(s).}
\end{table}

\subsection{Reference star spectra}

To ensure the consistency of our data between different observing runs we acquired spectra of asteroids Iris and/or Vesta, which are equivalent to solar spectra, and/or the bright solar twin star 18\,Sco (HIP\,79672) in each one of the runs (Table~\ref{t:reference}). Asteroids were observed if they were bright ($V\sim6.4-8.3$) during our runs; otherwise only 18\,Sco was observed in a given run. We acquired asteroid spectra in three runs and 18\,Sco in three runs as well, with only one of them in common with an asteroid observation. Two observations of 18\,Sco were made in different nights in one of the runs. Thus, there are four available 18\,Sco spectra. We did not produce solar or 18\,Sco spectra by co-adding data taken in different runs. Instead, we analyzed them independently.

\subsection{Absolute radial velocity} \label{s:rv}

Estimates of the absolute radial velocities (RVs) of our sample stars are required as a starting guess for the HARPS data reduction. We obtained those values using our MIKE extracted spectra as follows.

Twenty two of our solar twins are listed in the \cite{nidever02} catalog of RVs of stable stars (their Table~1). The RVs of these objects had been monitored for four years and they were found to be stable within 0.1\,\kms. The zero point of the \citeauthor{nidever02}\ RVs is consistent with the accurate RV scales of \cite{stefanik99} and \cite{udry99} within 0.1\,\kms, while their internal uncertainties are only about 0.03\,\kms. We used these 22 stars as RV standards.

The absolute radial velocities of our stars were determined by cross-correlation of their extracted spectra with the 22 RV standards mentioned above. We employed IRAF's \texttt{fxcor} task to perform the cross-correlations. Orders significantly affected by telluric absorption were excluded. The order-to-order relative RVs were averaged to get a single RV value. The 1\,$\sigma$ error of these averages ranges between 0.1 and 0.6\,\kms\ depending on the standard star used in the cross-correlation. Thus, for each program star, 22 RVs were measured, each corresponding to a different standard reference. These 22 RV values were weight-averaged to compute the final RV of each star. The average 1\,$\sigma$ error of these final averages is 0.6\,\kms, i.e., larger than the $1\,\sigma$ errors of the cross-correlations, suggesting that the uncertainties of the final averages are dominated by systematic errors in the RVs adopted for the standard stars and not by our observational noise. Table~\ref{t:sample} lists our derived absolute radial velocities.

Instead of adopting the \cite{nidever02} RVs for the standard stars, we re-derived their RVs with the same procedure described above, but using for each standard star the other 21 standards for cross-correlation. The average difference between the \cite{nidever02} RVs and those we re-derived for the 22 standard stars is $+0.01\pm0.58$\,\kms.

\subsection{Chromospheric activity}

We used the MIKE extracted spectra to calculate chromospheric activity indices for our stars. Naturally, the actual level of activity will be better determined using the time-series measurements of the HARPS spectra. At this stage, we are only interested in an initial reference estimate of these values.

First, we measured ``instrumental'' $S=(H+K)/(R+V)$ values using the \ion{Ca}{ii} H \& K fluxes ($H,K$) and their nearby continuum fluxes ($R,V$). The former were computed by flux integration using triangular filters of width=1.1\,\AA\ centered at 3933.7 ($K$) and 3968.5\,\AA\ ($H$). The (pseudo-)continuum fluxes were estimated as the flux averages at $3925\pm5\,\AA$ ($V$) and $3980\pm5\,\AA$ ($R$); as the regions are broad, spectral lines also fall in the pseudo-continuum. We used IRAF's \texttt{sband} task for these calculations. Then, we searched for standardized $S_\mathrm{MW}$ values (i.e., $S$ values on the Mount Wilson scale) previously published for our sample stars. We found $S_\mathrm{MW}$ values for 62 of our stars in the catalogs by \cite{duncan91,henry96,wright04,gray06,jenkins06,jenkins11} and \cite{cincunegui07}. Measurements of the same object found in more than one of these sources were averaged. We used these values to place our instrumental $S$ values into the Mount Wilson system via a second order polynomial fit to the $S$ vs.\ $S_\mathrm{MW}$ relation.

\begin{figure}
\centering
\includegraphics[bb=150 315 468 480,width=9.1cm]{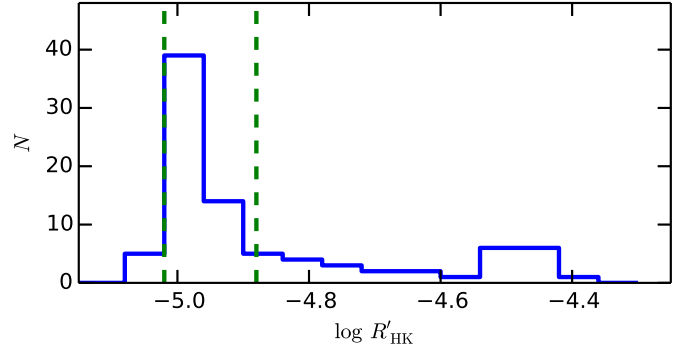}
\caption{Histogram of $\log R'_\mathrm{HK}$ values for our sample stars. The dashed lines limit the range of values covered by the 11-year solar cycle.}
\label{f:rhk}
\end{figure}

To calculate $\log R'_\mathrm{HK}$ values (given in Table~\ref{t:sample}) we employed the set of equations listed in Section~5.2 of \cite{wright04}, using our calibrated $S_\mathrm{MW}$ measurements and the stars' $(B-V)$ colors given in the {\it Hipparcos} catalog \citep{perryman97}. The average difference between our $\log R'_\mathrm{HK}$ values and those previously published in the papers mentioned in the paragraph above is $0.004\pm0.043$. We expect some of this scatter to be due to the intrinsic nature of stellar activity cycles. The solar cycle, for example, has a $\log R'_\mathrm{HK}$ span of about 0.1 \citep{hall09}. A histogram of our $\log R'_\mathrm{HK}$ values is given in Figure~\ref{f:rhk}.

As an independent check of our $\log R'_\mathrm{HK}$ calculations, we compared the values we derived with those computed by \cite{lovis11}, who used multi-epoch HARPS spectra. Fifteen of our sample stars are included in the study by \citeauthor{lovis11} They all show low levels of chromospheric activity ($\log R'_\mathrm{HK}\lesssim-4.9$). Compared to the mean values given in \citeauthor{lovis11}, our $\log R'_\mathrm{HK}$ values are, on average, only $0.005\pm0.025$ higher, i.e., in excellent agreement with theirs considering the calibration errors and potential activity cycle variations.

\section{Model atmosphere analysis}

A very important component of our work is the determination of iron abundances in the stars' atmospheres. To compute these values, we used the curve-of-growth method, employing the 2013 version of the spectrum synthesis code MOOG\footnote{\url{http://www.as.utexas.edu/~chris/moog.html}} \citep{sneden73} for the calculations (specifically the \texttt{abfind} driver). We adopted the ``standard composition'' MARCS grid of 1D-LTE model atmospheres \citep{gustafsson08},\footnote{\url{http://marcs.astro.uu.se}} and interpolated models linearly to the input $\teff,\logg,\feh$ values when necessary. As shown in previous works by our group, the particular choice of model atmospheres is inconsequential given the strict differential nature of our work and the fact that all stars are very similar to the Sun.

\subsection{Iron linelist}

A set of 91 \fei\ and 19 \feii\ lines were employed in this work (Table~\ref{t:linelist}). These lines were taken from our previous studies and most of them have transition probabilities measured in the laboratory. Nevertheless, the accuracy of these values and the fact that some lines have $\log gf$ values determined empirically are both irrelevant for our work. All these lines are in the linear part of the curve-of-growth in Sun-like stars and therefore their uncertainties cancel-out in a strict line-by-line differential analysis.

Most of our iron lines are completely unblended. However, our \fei\ linelist includes a few lines that are somewhat affected by other nearby spectral features. The reason to keep these lines is that they balance the excitation potential ($\chi$) vs.\ reduced equivalent width ($REW=\log EW/\lambda$, where $EW$ is the line's equivalent width) distribution of the \fei\ lines. This is important to avoid degeneracies and biases in the determination of stellar parameters using the standard excitation/ionization balance technique, which is described in Section~\ref{s:excion}.

\begin{figure}
\centering
\includegraphics[bb=110 210 510 595,width=9.1cm]{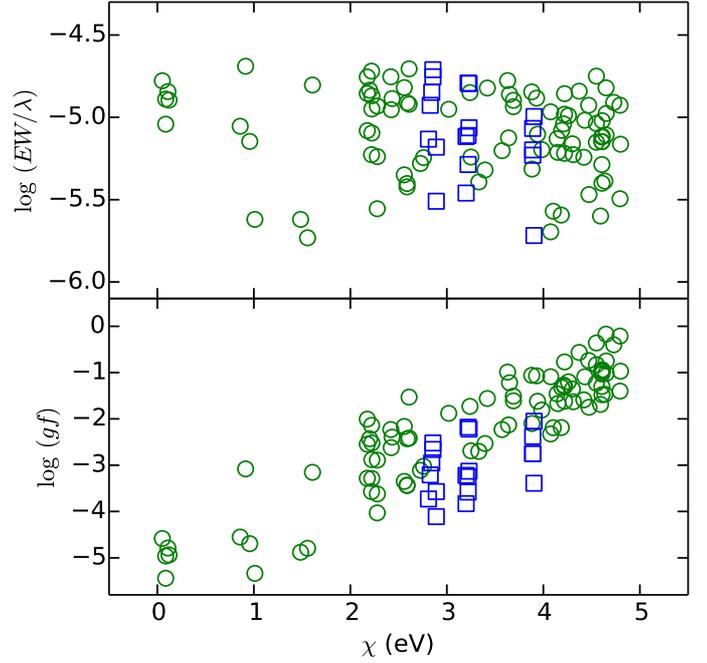}
\caption{Excitation potential versus reduced equivalent width (top panel) and transition probability (bottom panel) relations for our \fei\ (circles) and \feii\ lines (squares). The equivalent widths correspond to our solar reference, which is based on spectra of sunlight reflected from asteroids, as described in Section~\ref{s:ew}.}
\label{f:eprewgf}
\end{figure}

The $\chi$ vs.\ $REW$ relation of our iron linelist is shown in Figure~\ref{f:eprewgf}. In addition to retaining as many as possible low-$\chi$ lines, even if they are difficult to measure, we had to exclude a number of very good (i.e., clean) lines on the high-$\chi$ side, also to prevent biasing the stellar parameter determination. Having an unbalanced $\chi$ distribution would make the $\teff$ more sensitive to one particular type of spectral line, which should be avoided. The positive correlation between excitation potential and transition probability is expected. Lower $\chi$ lines tend to be stronger; to avoid saturated lines, lower $\log(gf)$ features are selected.

\subsection{Equivalent width measurements} \label{s:ew}

Equivalent widths were measured ``manually,'' on a star-by-star, line-by-line basis, using IRAF's \texttt{splot} task. Gaussian fits were preferred to reduce the impact of observational noise on the lines' wings, which has a stronger impact on Voigt profile fits. At the spectral resolution of our data, Gaussian fits are acceptable. The \texttt{deblend} option was used when necessary, making sure that the additional lines were consistently fitted for all stars (i.e., we used the same number and positions of blending lines). For some spectral lines a relatively low pseudo-continuum assessment was necessary due to the presence of very strong nearby lines. We also made sure to adopt consistent pseudo-continua for all stars. Our experience shows that manual measurement of $EW$s is superior to an automated procedure in terms of achieved consistency and accuracy.\footnote{Our team has employed a number of automated tools to calculate equivalent widths in the past. Although these procedures are extremely helpful when dealing with very large numbers of stars and long spectral line-lists, we have found that even minor issues with the continuum determination or other data reduction deficiencies always result in a small fraction of spectral lines with incorrect $EW$ values (or at least not precise enough when investigated in chemical abundance space). Sigma-clipping could be invoked to get rid of these outliers, but this introduces star-to-star inconsistencies in the derivation of stellar parameters. Despite being extremely inefficient, visual inspection of every spectral line and ``manual measurements'' have proven to be the most reliable and self-consistent techniques for $EW$ determination in our works.}

Given the characteristics of our data, the predicted error in our $EW$ measurements is about 0.2\,m\AA. This value was computed using the formula by \cite{cayrel88}, who points out that the true error is likely higher due to systematic uncertainties, for example those introduced by the continuum placement. The spectra of our reference stars were used to get a better estimate of our $EW$ errors.

\begin{figure}
\includegraphics[bb=110 255 500 550,width=9cm]{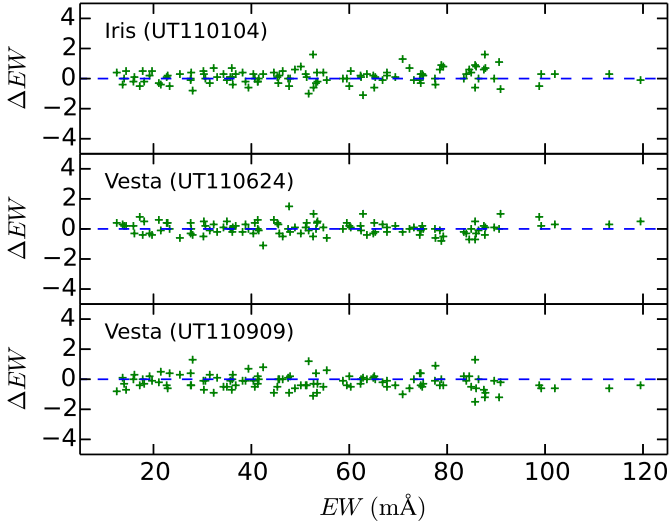}
\caption{Equivalent width differences for our asteroid (solar) spectra. In each panel, the differences between the $EW$s measured in a given spectrum and the average $EW$ values are shown. On the top left side of each panel, the name of the target is followed by the UT date of observation (in YYMMDD format).}
\label{f:ew_sun}
\end{figure}

\begin{figure}
\includegraphics[bb=110 210 500 590,width=9cm]{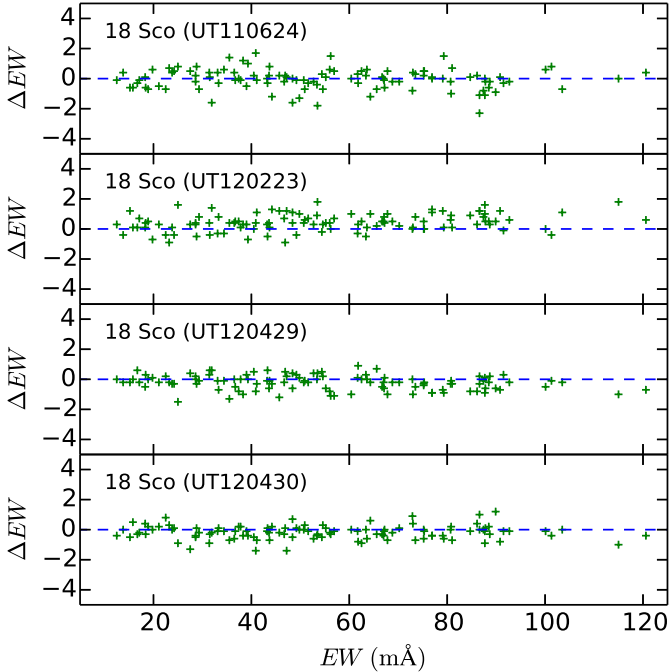}
\caption{As in Figure~\ref{f:ew_sun} for our 18\,Sco spectra.}
\label{f:ew_18sco}
\end{figure}

The $EW$s measured in our three solar (asteroid) spectra were averaged to create our adopted solar $EW$ list, which is provided in Table~\ref{t:linelist}. The error bars listed there for $EW_\odot$ correspond to the $1\,\sigma$ scatter of the three equivalent width measurements available for each spectral line. Similarly, we averaged the $EW$s of the four spectra of 18\,Sco. The difference between these average $EW$s and those measured in the individual spectra are shown in Figures~\ref{f:ew_sun} and \ref{f:ew_18sco}. The 1\,$\sigma$ scatter of the $EW$ differences shown in Figures~\ref{f:ew_sun} and \ref{f:ew_18sco} ranges from 0.4 to 0.5\,m\AA\ and from 0.5 to 0.7\,m\AA, respectively.

Assuming that the scatter values quoted above arise from purely statistical errors, the $EW$ uncertainty for the 18\,Sco spectra can be estimated as $\sqrt{3/4}\times(0.5-0.7)$.\footnote{If the $EW$ precision of each spectrum is $\sigma$, that of the average of $n$ has a precision of $\sigma_n=\sigma/\sqrt{n-1}$. The scatter of the average minus individual spectrum measurement differences is then $\sqrt{\sigma^2+\sigma_n^2}=\sigma\sqrt{n/(n-1)}$. Thus, $\sigma$ is proportional to $\sqrt{(n-1)/n}$, with $n=4$ in our 18\,Sco example.} Since the individual 18\,Sco spectra are typical of all our other sample star observations, we estimate an average $EW$ uncertainty of about 0.5\,m\AA, which corresponds to 1\,\% for a line of $EW=50\,$m\AA.

The good agreement found in our $EW$ measurements made on multiple spectra for the same stars (Sun and 18\,Sco) ensures a high degree of consistency in the derived relative stellar parameters, as will be shown quantitatively in Section~\ref{s:18sco}.

\subsection{Spectroscopic parameters} \label{s:excion}

We employed the excitation/ionization balance technique to find the stellar parameters that produce consistent iron abundances. We started with literature values for the stars' fundamental atmospheric parameters $\teff,\logg,\feh,\vt$ and iteratively modified them until the correlations with $\chi$ and $REW$ were minimized, while simultaneously minimizing also the difference between the mean iron abundances derived from \fei\ and \feii\ lines separately.\footnote{A Python package (qoyllur-quipu, or $q^2$) has been developed by I.R.\ to simplify the manipulation of MOOG's input and output files as well as the iterative procedures. The $q^2$ source code is available online at \url{https://github.com/astroChasqui/q2}.} The stellar parameters derived in this manner are often referred to as ``spectroscopic parameters.''

We used a strict differential approach for the calculations described here. This means that the stars' iron abundances were measured relative to the solar iron abundance on a line-by-line basis. Thus, if $A_{\mathrm{Fe},i}$ is the absolute iron abundance derived for a spectral line $i$, the following quantities: $\feh_i=A_{\mathrm{Fe},i}-A_{\mathrm{Fe},i}^\odot$ were employed to perform the statistics and to calculate the final relative iron abundances by averaging them. Strict differential analysis minimizes the impact of model uncertainties as well as errors in atomic data because they cancel-out in each line calculation. This is particularly the case when the sample stars are all very similar to each other and very similar to the star employed as reference, i.e., the Sun in our case. We adopted $\teff^\odot=5777$\,K, $\logg^\odot=4.437$, $\vt^\odot=1.0$\,\kms, and the absolute solar abundances by \cite{anders89}. The particular choice of the latter has no effect on the precision of our relative abundances.

In each iteration, we examined the slopes of the $\feh$ vs.\ $\chi$ and $\feh$ vs.\ $REW$ relations. If they were found positive (negative), the $\teff$ and $\vt$ values were increased (decreased). At the same time, if the mean \fei\ minus \feii\ iron abundance difference was found positive (negative), the $\logg$ value was increased (decreased). We stopped iterating when the standard deviations of the parameters from the last five iterations were all lower than 0.8 times the size of the variation step. The first set of iterations was done with relatively large steps; the $\teff$, $\logg$, and $\vt$ parameters were modified by $\pm32$\,K, $\pm0.32$, and $\pm$0.32\,\kms, respectively. After the first convergence, the steps were reduced in half, i.e., to $\pm16$\,K, $\pm$0.16, and $\pm$0.16\,\kms, and so on, until the last iteration block, in which the steps were $\pm$1\,K, $\pm$0.01, and $\pm$0.01\,\kms.

The average $\feh$ resulting in each iteration was computed, but not forced to be consistent with the input value. While this condition should be enforced by principle, within our scheme one could save a significant amount of computing time by avoiding it. After all, the final iteration loop has such small steps that the input and resulting $\feh$ values will not be significantly different. Indeed, the average difference between input and output $\feh$ values from all last iterations in our work is $0.0002\pm0.0025$\,dex.

The average line-to-line scatter of our stars' derived $\feh$ values is 0.02\,dex (including both \fei\ and \feii\ lines). Formal errors for the stellar parameters $\teff$, $\logg$, and $\vt$ were computed as in \cite{epstein10} and \cite{bensby14}. On average these formal errors are $\sigma(\teff)=7$\,K, $\sigma(\logg)=0.019$, and $\sigma(\vt)=0.016$\,\kms. For $\feh$, the formal error was computed by propagating the errors in the other atmospheric parameters into the $\feh$ calculation; adding them in quadrature (therefore assuming optimistically that they are uncorrelated) and including the standard error of the mean line-to-line $\feh$ abundance. On average, $\sigma(\feh)=0.006$\,dex. The stellar parameter uncertainties are the main source of this error, not the line-to-line scatter.

The errors quoted above are extremely low due to the high quality of our data and our precise, consistent, and very careful $EW$ measurements. One should keep in mind, however, that the true meaning of these formal errors is the following: inside their range, the $\feh$ versus $\chi$/$REW$ slopes and \fei\ minus \feii\ iron abundance differences are consistent with zero within the $1\,\sigma$ line-to-line scatter. In other words, they just correspond to the precision with which we are able to minimize the slopes and iron abundance difference. Rarely do they represent the true errors of the atmospheric parameters because they are instead largely dominated by systematic uncertainties \cite[e.g.,][]{asplund05:review}. The only possible exceptions, as argued before, are solar twin stars if analyzed relative to the Sun or relative to each other.

\begin{figure}
\centering
\includegraphics[bb=155 190 468 605,width=9.1cm]{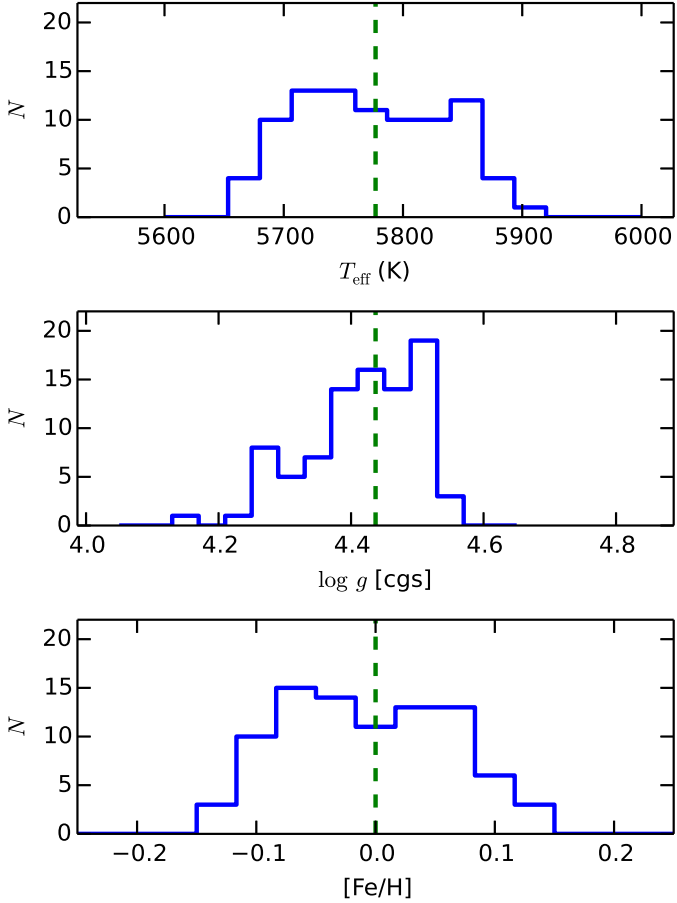}
\caption{Histograms of stellar parameters for our sample stars. The dashed lines correspond to the canonical solar values.}
\label{f:pars}
\end{figure}

Our derived spectroscopic parameters, and their internal errors, are given in Table~\ref{t:pars}. Figure~\ref{f:pars} shows our sample histograms for these stellar parameters. The reliability of our error estimates and detailed accuracy assessments are discussed later in this paper. For now, it is interesting to compare our spectroscopic parameters with those determined by \cite{sousa08}, who employed essentially the same technique used in this work, but with some differences regarding the ingredients of the process. Nineteen stars were found in common between our work and the study by \citeauthor{sousa08}. The average differences in stellar parameters, in the sense \citeauthor{sousa08} minus this work, are: $\Delta\teff=-5\pm12$\,K (5660 to 5875\,K), $\Delta\logg=0.01\pm0.04$ (4.16 to 4.49), and $\Delta\feh=-0.004\pm0.015$ (--0.13 to +0.12). The ranges in parenthesis correspond to those of the subsample of stars in common between the two studies.

\subsection{18\,Sco as a test case} \label{s:18sco}

\begin{figure}
\centering
\includegraphics[bb=90 115 520 690,width=9.1cm]{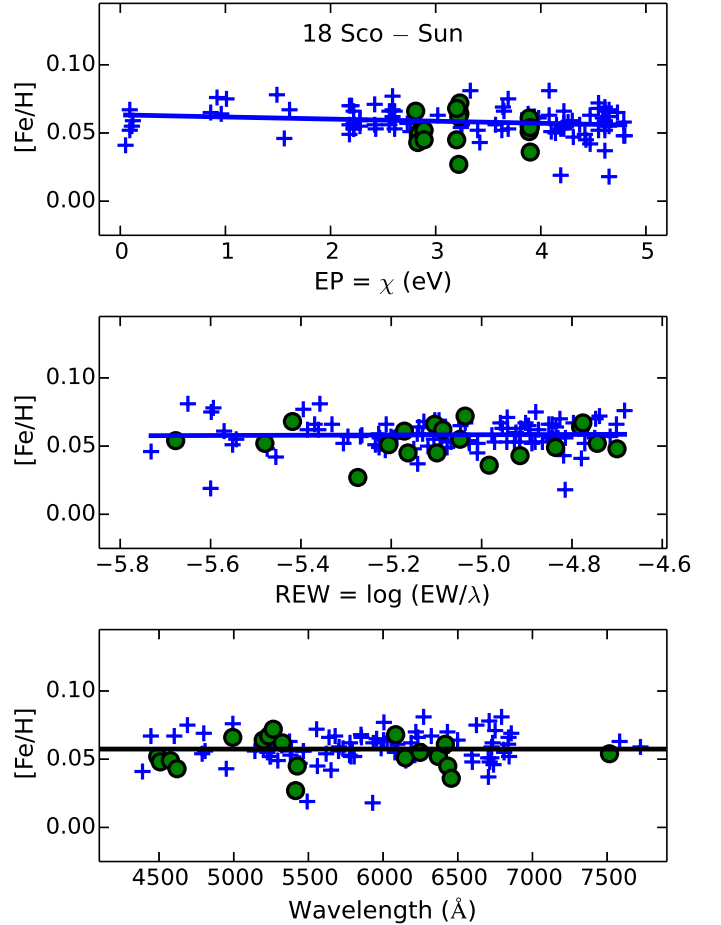}
\caption{Line-to-line relative iron abundance of 18\,Sco as a function of excitation potential (top panel), reduced equivalent width (middle panel), and wavelength (bottom panel). Crosses (circles) are \fei\ (\feii) lines. The solid lines in the top and middle panels are linear fits to the \fei\ data. In the bottom panel, the solid line is a constant which corresponds to the average iron abundance of this star.}
\label{f:18Sco}
\end{figure}

Figure~\ref{f:18Sco} shows an example of a final, fully converged solution. It corresponds to the ``closest-ever'' \citep{porto97}, bright solar twin star 18\,Sco. Since the $EW$ values employed in this calculation for 18\,Sco correspond to the average of four independent observations, each made with a $S/N\sim400$ spectrum, this is our most precise case: $\teff=5814\pm3$\,K, $\logg=4.45\pm0.01$, $\feh=0.056\pm0.004$, and $\vt=1.02\pm0.01$\,\kms. The precision of our results for all other stars is typically half as good, yet still extremely precise.

As a test case, we determined the stellar parameters of 18\,Sco using each of its four available individual spectra. These are more representative of our sample stars' data in general. The parameters derived from these individual spectra have the following mean and $1\,\sigma$ values: $\teff=5816\pm4$\,K, $\logg=4.445\pm0.005$, $\feh=0.053\pm0.003$, and $\vt=1.025\pm0.015$\,\kms. This is fully consistent with the values {\it and errors} derived for the case when the average $EW$ values of 18\,Sco are employed, which ensures that our formal error calculation is reliable.

We performed a similar test calculation for the three available solar spectra relative to their average $EW$s. We found the following averages for the three individual spectra: $\teff=5778\pm3$\,K, $\logg=4.433\pm0.012$, $\feh=-0.001\pm0.006$, and $\vt=1.000\pm0.014$\,\kms, which further demonstrates that the formal errors we derived fully correspond to the observational noise.

\cite{melendez14:18sco} have recently used spectra of 18\,Sco taken with the UVES and HIRES spectrographs on the VLT and Keck Telescopes, respectively, to determine highly precise parameters of this star. The reference solar spectra in their study are reflected sunlight observations from the asteroids Juno (for the VLT case) and Ceres (for HIRES). The parameters found for 18\,Sco in that study are $\teff=5823\pm6$\,K, $\logg=4.45\pm0.02$, and $\feh=+0.054\pm0.005$. All these values are consistent with those derived with our MIKE spectra within the $1\,\sigma$ precision errors. For $\teff$, note that the $1\,\sigma$ lower limit of the value from \cite{melendez14:18sco} is exactly the same as the $1\,\sigma$ upper limit from our work (our most precise value, that from the average $EW$ measurements, is $\teff=5814\pm3$\,K).

\subsection{Isochrone masses and ages} \label{s:massandage}

The most common approach to derive stellar masses and ages of large samples of single field stars is the isochrone method \cite[e.g.,][]{lachaume99}. In most implementations, this method uses as inputs the observed $\teff$, $M_V$ (absolute magnitude), and $\feh$ values, along with their errors. To calculate $M_V$, a measurement of the star's parallax is required. The latter is available in the {\it Hipparcos} catalog \citep{vanleeuwen07} for most Sun-like stars in the solar neighborhood.

Each data point in an isochrone grid has stellar parameters associated to it, including $\teff$, $M_V$, and $\feh$, but also mass, radius, age, luminosity, etc. Thus, one could find the isochrone point with $\teff$, $M_V$, and $\feh$ closest to the observed values and associate the other stellar parameters to that particular observation. To achieve higher accuracy, one can calculate a probability distribution for each of the unknown parameters using as weights the distances between observed and isochrone $\teff$, $M_V$, and $\feh$ values (normalized by their errors). Then, the probability distributions can be employed to calculate the most likely parameter values and their formal uncertainties.

Isochrone age determinations are subject to a number of sampling biases whose impact can be minimized using Bayesian statistics \cite[e.g.,][]{pont04,jorgensen05,dasilva06,casagrande11}. While correcting for these biases is crucial for statistical stellar population studies, their importance is secondary for small samples of field stars spanning a narrow range in parameters where the main goal is to sort stars chronologically \cite[e.g.,][]{baumann10}. Because of the small internal errors of our observed stellar parameters, this is possible with our approach.

To derive a precise stellar parameter using isochrones, the star must be located in a region of stellar parameter space where the parameter varies quickly along the evolutionary path. In particular, a precise age can be calculated for stars near the main-sequence turn-off. On the main-sequence, isochrones of different ages are so close to each other that a typical observation cannot be used to disentangle the isochrone points that correspond to that star's age. It is for this reason that it is often assumed that isochrone ages of main-sequence stars are impossible to calculate.

Nevertheless, solar twin stars offer the possibility of deriving reasonably precise isochrone ages, at least on a relative sense. The most important step to achieve this goal consists in replacing $M_V$ for $\logg$ as one of the input parameters. The latter can be derived with extremely high precision, as we have done in Section~\ref{s:excion}. Thus, even though main-sequence isochrones are close to each other in the $\teff$ versus $\logg$ plane, the high precision of the observed $\teff$ and $\logg$ values ensures that the age range of the isochrones that are consistent with these high-quality observations is not too wide. For example, \cite{melendez12} showed that the isochrone age of ``the best solar twin star'' HIP\,56948 can only be said to be younger than about 8\,Gyr if its {\it Hipparcos} parallax is employed to calculate $M_V$, which is in turn used as input parameter, but constrained to the 2.3--4.1\,Gyr age range if its very precise spectroscopic $\logg$ value is adopted as input parameter instead (see their Section 4.3 and Figure 10).

\begin{figure}
\centering
\includegraphics[bb=85 265 530 530,width=9.0cm]{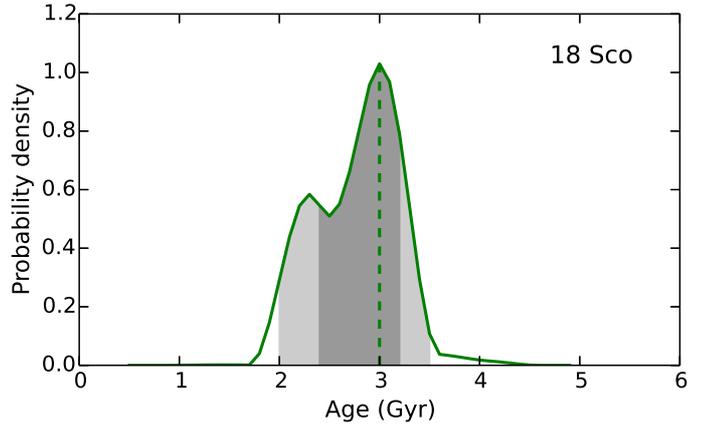}
\caption{Age probability distribution of 18\,Sco. The dashed line is at the most probable age. The probability density units are arbitrary. The dark (light) gray shaded area corresponds to the $1\sigma$ ($2\sigma$) confidence interval.}
\label{f:age}
\end{figure}

Isochrone masses and ages depend on the grid employed. This is true regardless of which input parameters are used. However, this leads mostly to systematic errors and, by definition, does not affect the internal precision of the parameters derived.

\begin{figure}
\centering
\includegraphics[bb=150 255 468 547,width=9.1cm]{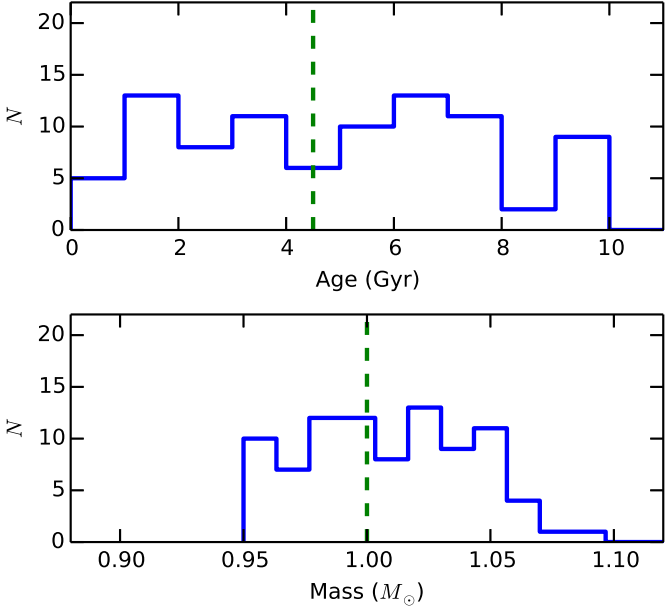}
\caption{Histograms of stellar mass and age for our sample stars. The dashed lines correspond to the canonical solar values.}
\label{f:mass_age}
\end{figure}

In this work, we employed the isochrone method implementation by \cite{ramirez13:thin-thick}, but adopting the spectroscopic $\logg$ instead of $M_V$ as input parameter. \cite{ramirez13:thin-thick} implementation uses the Yonsei-Yale isochrone set \cite[e.g.,][]{yi01,kim02}. Figure~\ref{f:age} shows the age probability distribution of 18\,Sco as an example. The asymmetry of this curve is a common feature of isochrone age probability distributions. We assign the most probable value from this distribution as the age of the star. Confidence intervals at the 68\,\% and 96\,\% levels can then be interpreted as the $1\,\sigma$ and $2\,\sigma$ limits of the star's age. They are represented by the dark- and light-gray shaded areas in Figure~\ref{f:age}. Mass probability distributions are nearly symmetric; thus a single value is sufficient to represent its internal uncertainty. The derived isochrone masses and ages ($\tau$) for our sample stars are given in Table~\ref{t:pars}. Sample histograms for these parameters are shown in Figure~\ref{f:mass_age}.

The internal precision of our ages varies widely from star-to-star. On average, the $1\,\sigma$ age range from the probability distributions is 1.3\,Gyr, with a maximum value of 3.1\,Gyr and a minimum at 0.5\,Gyr. It must be stressed that these numbers should not be quoted as the absolute age uncertainties from our work. They only represent the internal precision with which we are able to find nearby Yonsei-Yale isochrone points. For our present purposes, this is certainly acceptable, as we are mainly interested in the chronology of the sample and not necessarily the star's true ages (i.e., relative ages rather than absolute ages).

\begin{figure}
\centering
\includegraphics[bb=110 240 510 565,width=9.1cm]{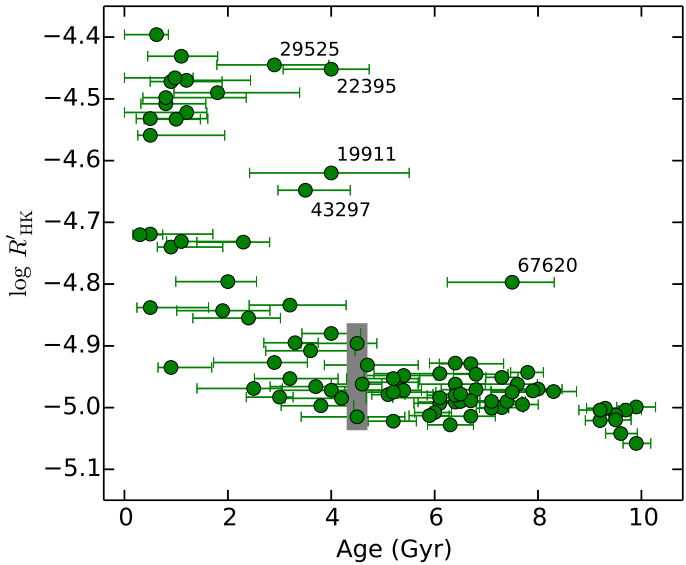}
\caption{Stellar age versus chromospheric activity index. Outlier stars are labeled with their HIP numbers. The gray bar at 4.5\,Gyr represents the range covered by the 11-year solar cycle.}
\label{f:age_rhk}
\end{figure}

To test the internal precision of our derived stellar ages, we plot them against our measured chromospheric activity indices $\log R'_\mathrm{HK}$ in Figure~\ref{f:age_rhk}. Stellar activity is predicted to decay with time due to rotational braking \cite[e.g.,][]{skumanich72,barnes10}, but plots of activity index versus stellar age typically have very large scatter. The most likely explanation for this is that the activity indices depend on the stars' effective temperatures. In our case, the resulting relation is very tight (after excluding the outliers; see below). This owes to the fact that these objects are all very similar one-solar-mass, solar-metallicity main-sequence stars.

There are a few outliers in the $\log R'_\mathrm{HK}$ versus age plot of Figure~\ref{f:age_rhk}. For example, one star at age=7.5\,Gyr has a $\log R'_\mathrm{HK}$ value about 0.15 above that of all other coeval stars. This object, HIP\,67620, is known to have an unresolved (for our spectroscopic observations) faint companion, which is revealed by speckle interferometry \citep{hartkopf12}. The other outliers are also unusually high activity stars, but in the age range between about 3 and 4\,Gyr; they are HIP\,19911, HIP\,22395, HIP\,29525, and HIP\,43297.

Figure~\ref{f:age_rhk} shows an overall decrease of stellar activity with increasing age. The star-to-star scatter at a given age is most likely dominated by intrinsic changes in the activity of the stars and not due to observational errors. During the 11-year solar cycle, for example, the $\log R'_\mathrm{HK}$ index of the Sun varies from about $-5.02$ to $-4.88$ \citep{hall09}. This range is illustrated by the gray bar at age 4.5\,Gyr in Figure~\ref{f:age_rhk}. The solar twins of age close to solar span a $\log R'_\mathrm{HK}$ range compatible with the solar data, suggesting that the activity levels of the present-day solar cycle are typical of other Sun-like stars of solar age.

The larger scatter seen in younger stars suggests that they have larger variations in their activity levels. On the other end, the very low $\log R'_\mathrm{HK}$ values of the oldest solar twins ($\mathrm{age}\gtrsim9$\,Gyr) suggest that the mean chromospheric activity continues to decrease, albeit slowly, as stars get significantly older.

The exact nature of the activity--age relation of solar twin stars will be more clear once the multi-epoch HARPS data are analyzed. Even though four years is not enough to cover the solar cycle, younger one-solar-mass stars may have shorter cycles, which could allow us to measure their full $\log R'_\mathrm{HK}$ ranges. For example, 18\,Sco has a cycle of about 7\,years \citep{hall07}. We will investigate thoroughly the age dependency of stellar activity of solar twins in a future publication.

\section{Validation}

The fundamental atmospheric parameters $\teff$ and $\logg$ can be determined using techniques which are independent of the iron line (spectroscopic) analysis described in the previous section. To be more precise, these alternative techniques are {\it less dependent} on the iron line analysis; the average $\feh$ could still play a role as input parameter.

As will be described below, $\teff$ can be measured using the star's photometric data or $\ha$ line profile. Also, $\logg$ can be estimated using a direct measurement of the star's absolute magnitude, which requires a knowledge of its trigonometric parallax. The $\feh$ value derived for each spectral line depends on the input $\teff$ and $\logg$. Thus, if they are different from the ``spectroscopic'' values derived before, the slopes of the $\feh$ versus $\chi$ and $REW$ relations will no longer be zero. The same will be true for the mean \fei\ minus \feii\ iron abundance difference.

Determining $\teff$ and $\logg$ using other methods requires relaxing the conditions of excitation and ionization equilibrium. However, one could derive a consistent $\vt$ value by forcing the $\feh$ versus $REW$ slope to be zero. The resulting average $\feh$ value will be different than the spectroscopic one, which may in turn have an impact on the alternative $\teff$ and $\logg$ values. Therefore, strictly speaking, one must iterate until all parameters are internally consistent.

Finding consistent solutions for these alternative parameters would be necessary if we were interested in using them in our work. As will be shown later, these parameters are not as internally precise as those we inferred from the spectroscopic analysis of the previous section. We will thus only use the latter in our future works. In this section, we are only interested in calculating these parameters given one different ingredient. In that way, we can better understand the sources of any important discrepancies. Therefore, in deriving $\teff$ or $\logg$ using other methods, we kept all other parameters constant and no attempt was made to achieve self-consistent results using iterative procedures.

\subsection{IRFM effective temperatures}

One of the most reliable techniques for measuring with accuracy a solar-type star's effective temperature is the so-called infrared flux method (IRFM). First introduced by \cite{blackwell79}, the IRFM uses as $\teff$ indicator the ratio of monochromatic (infrared) to bolometric flux, which is independent on the star's angular diameter. Observations of that ratio based on absolutely calibrated photometry are compared to model atmosphere predictions to determine $\teff$. The flux ratio is highly sensitive to $\teff$ and weakly dependent on $\logg$ or $\feh$. In addition, systematic uncertainties due to model simplifications are less important in the infrared. Modern implementations of the IRFM \cite[e.g.,][]{alonso96:irfm,ramirez05a,gonzalez-hernandez09,casagrande10} have been very valuable for a large number of investigations in stellar astrophysics.

In this work, we used the implementation described in \cite{casagrande10}, which uses multiband optical (Johnson-Cousins) and infrared (2MASS) photometry to reconstruct the monochromatic (infrared) and bolometric flux. The accuracy of such implementation has been tested thoroughly (see \citealt{casagrande14} for a summary). We used the homogeneous set of $BV(RI)_C$ and $JHK_S$ photometry published in our earlier investigations of Sun-like stars \citep{ramirez12_suncolor,casagrande12}. The overlap with this study is restricted to 30 objects. The flux outside photometric bands is estimated using the \cite{castelli04} theoretical model fluxes interpolated at the spectroscopic $\feh$, $\logg$, and (for only the first iteration) $\teff$ values of each star.

While the IRFM is only mildly sensitive to the adopted $\feh$ and $\logg$, the spectroscopic $\teff$ is used only as an input parameter, and an iterative procedure is adopted to converge in $\teff$(IRFM). We checked that convergence is reached independently on the input $\teff$, be it spectroscopic or a random point in the grid of synthetic fluxes. Errors in $\teff$(IRFM) were computed as the sum in quadrature of the scatter in the three $JHK_S$ $\teff$ values and the error due to photometric errors, computed using Monte Carlo experiments. A constant value of 20\,K was added linearly to these errors to account for the uncertainty in the zero point of the IRFM $\teff$ scale. \cite{casagrande10} showed that IRFM and direct, i.e., interferometric, $\teff$ values are offset by 18\,K.

\begin{figure}
\centering
\includegraphics[bb=110 210 510 595,width=9.1cm]{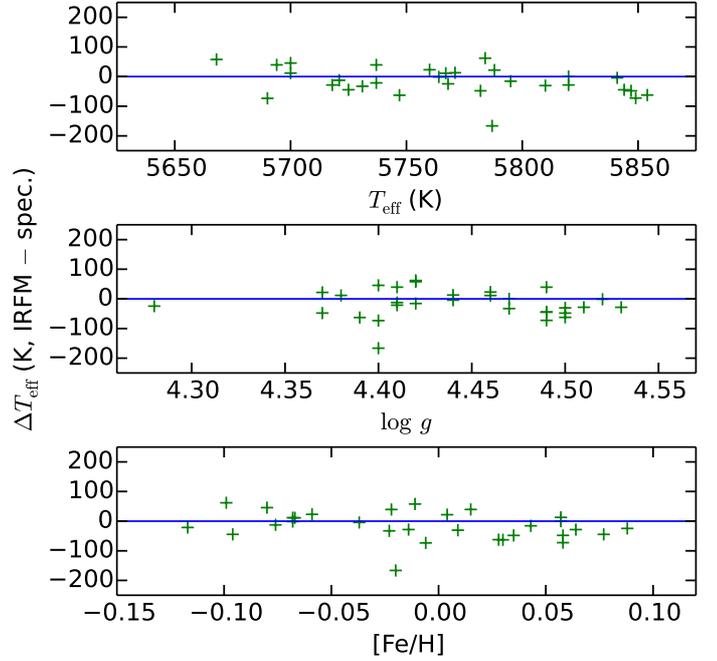}
\caption{IRFM minus spectroscopic $\teff$ as a function of spectroscopic parameters. The solid line is at zero.}
\label{f:teff_irfm}
\end{figure}

The differences between IRFM and spectroscopic effective temperatures for the 30 stars in our sample with homogeneous and accurate photometric data are shown in Figure~\ref{f:teff_irfm} as a function of atmospheric parameters. There are no significant offsets or correlations. On average, $\Delta\teff=-17\pm47$\,K (IRFM minus spectroscopic). The mean value of the errors in $\teff$(IRFM) is 52\,K, which shows that the scatter in Figure~\ref{f:teff_irfm} is fully explained by the IRFM uncertainties.

It is important to re-emphasize that the IRFM $\teff$ scale of \cite{casagrande10} has been thoroughly tested, in particular its absolute calibration. It has been shown to have a zero point that is accurately and precisely consistent with that which corresponds to the most reliable effective temperatures available for restricted samples of calibrating stars, for example those with accurate measurements of angular diameter or spectrophotometry. The fact that the spectroscopic effective temperatures of a representative group from our solar twin sample are on the same level as the most reliable IRFM $\teff$ values ensures that the $\teff$ scale adopted in this work has a zero point consistent with the best direct $\teff$ determinations.

\subsection{Effective temperatures from $\ha$ line-wing analysis}

The wings of Balmer lines in cool dwarf stars have been shown to be highly sensitive to the effective temperature while showing only a mild dependency on other stellar parameters such as $\logg$ or $\feh$ \citep[e.g.,][]{gehren81,fuhrmann93,barklem02}. Since they form in deep layers of the stars' atmospheres, their modeling is expected to be largely insensitive to non-LTE effects (see, however, \citealt{barklem07}), but dependent on the details of the treatment of convection \cite[e.g.,][]{ludwig09:ha}. These potential systematic uncertainties in the modeling of the Balmer lines will affect our sample stars in a very similar manner, which implies that we can determine a set of internally precise $\teff(\ha)$ values. We restrict our work to the Balmer $\ha$ line because it is the least affected by overlapping atomic features and it is the most amenable to proper continuum normalization (see below).

The $\ha$ line is very wide and it occupies nearly one-half of one of the orders in our MIKE extracted spectra. The normalization procedure described in Section~\ref{s:observations} does not result in a properly normalized $\ha$ profile because the $\ha$ line wings are not correctly disentangled from the local continuum. For echelle spectra, a better continuum normalization can be done by interpolating the shape of the continua and blaze functions of nearby orders to the order containing the $\ha$ line. The details of this procedure are described in \cite{barklem02}. In this work, we employed the six orders nearest to the $\ha$ order and normalized them using 10th order polynomials (only the half of each order that aligns with the position of the $\ha$ line was used). Then, we fitted a 3rd order polynomial to the order-to-order continuum data for each pixel along the spatial axis. The value of these polynomials at the order where $\ha$ resides was then adopted as the continuum for that order.

To derive $\teff(\ha)$ we employed model fits to the observed $\ha$ lines using $\chi^2$ minimization. We adopted the theoretical grid of $\ha$ lines by \cite{barklem02} for this procedure. Since real spectra of Sun-like stars contain many weak atomic features on top of the $\ha$ line, the $\chi^2$ was computed only in those small spectral windows free from weak line contamination. The latter have to exclude also the telluric lines present in our spectra. Since their position changes from star to star due to the differences in radial velocity and epoch of observation, these clean spectral windows are different for each spectrum. The internal precision of our $\teff(\ha)$ values was estimated from the $\teff$ versus $\chi^2$ relation as follows: $\sigma(\teff)=[2/(\partial^2\chi^2/\partial\teff^2)]^{1/2}$.

We applied the technique described above to calculate the solar $\teff(\ha)$. The average value from our three solar (asteroid) spectra is $5731\pm21$\,K. Even though this value is inconsistent with the nominal $\teff^\odot=5777$\,K, it is in excellent agreement with the solar $\teff(\ha)$ derived by \cite{barklem02}, who employed the very high quality ($R\gtrsim500\,000$, $S/N\gtrsim1\,000$) solar spectrum by \cite{kurucz84}. This ensures that our continuum normalization for the solar spectrum was performed correctly. The discrepancy for the solar effective temperature is intrinsic to the adopted 1D-LTE modeling of the $\ha$ line and not due to observational errors \citep{pereira13}. Given the similarity of our sample stars, and the fact that we are exploiting differential analysis, we can apply a constant offset of +46\,K to all our $\teff(\ha)$ as a first order correction.

The +46\,K offset leads to a $\teff^\odot(\ha)=5777$\,K. Interestingly, for 18\,Sco we find $\teff(\ha)=5772\pm18$\,K (average of the four 18\,Sco spectra available). After applying the +46\,K correction, this value becomes $\teff(\ha)=5818$\,K, which is in excellent agreement with our spectroscopic temperature for this star ($\teff=5814$\,K).

\begin{figure}
\centering
\includegraphics[bb=110 190 515 610,width=9.2cm]{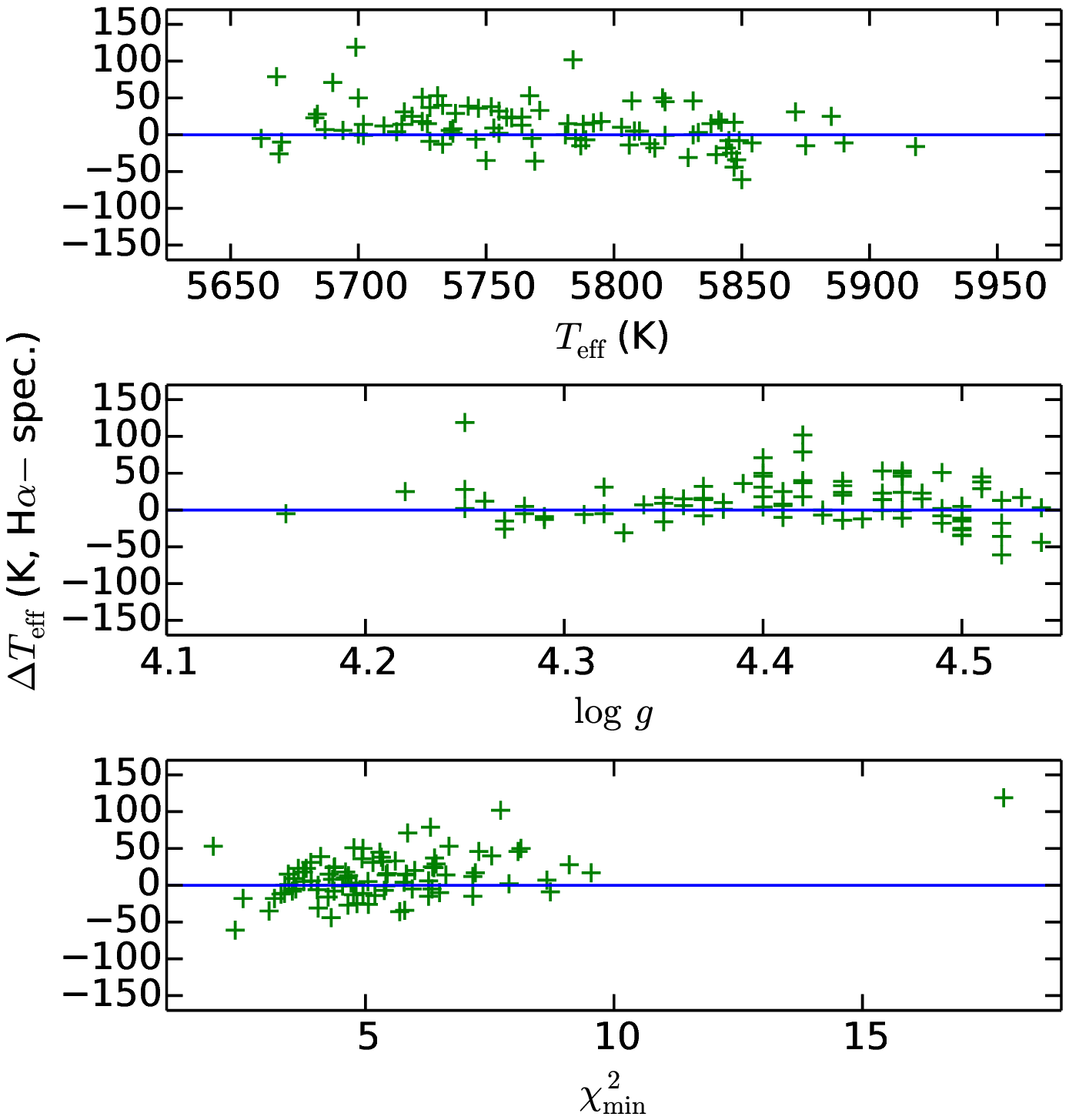}
\caption{$\ha$ minus spectroscopic $\teff$ as a function of $\teff$, $\logg$, and $\chi^2$ of the best $\teff(\ha)$ model fit. The solid line is at zero.}
\label{f:teff_halpha}
\end{figure}

Figure~\ref{f:teff_halpha} shows the differences between $\ha$ and spectroscopic $\teff$ for our entire sample. The average difference ($\ha$ minus spectroscopic) is $\Delta\teff=12\pm30$\,K (after applying the +46\,K offset to the $\ha$ temperatures of all stars). The average internal $\teff$ errors are only 23\,K and 7\,K for $\ha$ and spectroscopic $\teff$, respectively. Thus, the expected scatter for these differences is 24\,K, assuming no systematic trends, which do appear to exist. Figure~\ref{f:teff_halpha} reveals a small trend with $\teff$ such that $\Delta\teff$ seems to be slightly more positive for the cooler solar twins. These stars have stronger contaminant lines, which may lower the level of the $\ha$ line wing regions, leading to higher $\teff(\ha)$ values. The complex $\Delta\teff$ versus $\logg$ trend is difficult to understand, but it is clear that the agreement in $\teff$ values is excellent for solar twins of $\logg<4.4$.

The bottom panel of Figure~\ref{f:teff_halpha} shows $\Delta\teff$ as a function of $\chi^2_\mathrm{min}$, i.e., the $\chi^2$ value of the best-fit model. Lower $\chi^2_\mathrm{min}$ values imply a much better overall agreement between model and observations. Systematically high $\teff(\ha)$ values are found for $\chi^2_\mathrm{min}>5$ stars. Indeed, excluding them we find an average difference of $\Delta\teff=3\pm25$\,K. This star-to-star scatter is in excellent agreement with the expected one if systematic trends do not exist. Although we were very careful in our continuum determination and the selection of clean spectral windows for the $\chi^2$ measurements, the spectra with $\chi^2_\mathrm{min}>5$ were probably those in which the continuum normalization did not work as well (or precisely in the same exact fashion) as in the case of the solar spectrum. Note in particular that the largest $\Delta\teff$ occurs in the star with the worst $\chi^2_\mathrm{min}$.

Small inconsistencies in the continuum determination, which is already challenging for the $\ha$ line, and the impact of weak atomic lines contaminating the $\ha$ line wings are most likely responsible for the barely noticeable differences between $\ha$ and spectroscopic $\teff$ values of our solar twins. When we restrict the comparison to those stars with the best $\ha$ line normalization, the average difference (and star-to-star scatter) is perfectly consistent with zero within the expected internal errors.

\subsection{Trigonometric surface gravities}

For nearby stars, the trigonometric parallaxes from {\it Hipparcos} can be employed to calculate their absolute visual magnitudes with high precision. The well-defined location of the stars on the $M_V$ versus $\teff$ plane can then be used to calculate the stars' parameters by comparison with theoretical isochrones, as described in Section~\ref{s:massandage}.

To calculate absolute magnitudes we used previously measured apparent magnitudes and trigonometric parallaxes. The parallaxes we used are from the new reduction of the {\it Hipparcos} data by \cite{vanleeuwen07}. Visual magnitudes were compiled from various sources. First, we searched in the catalog by \cite{ramirez12_suncolor}, which is the most recent and comprehensive homogeneous $UBV(RI)_\mathrm{C}$ photometric dataset for solar twin stars. If not available in that catalog, we searched for Johnson's $V$ magnitudes in the General Catalog of Photometric Data (GCPD) by \cite{mermilliod97}. Then, we looked for ground-based $V$ magnitudes listed in the {\it Hipparcos} catalog (i.e., not the transformed $V_T$ magnitudes, but previous ground measurements of visual magnitudes compiled by the {\it Hipparcos} team). For a few stars we employed the $V$ magnitudes listed in the Str\"omgren catalogs of the GCDP or $V$ magnitudes calculated from {Hipparcos}' $V_T$ values. We used the $V$ magnitude errors reported in each of these sources, if available. Otherwise, we adopted the average of the errors reported, which is 0.012\,mag.

\begin{figure}
\centering
\includegraphics[bb=115 240 500 565,width=9.1cm]{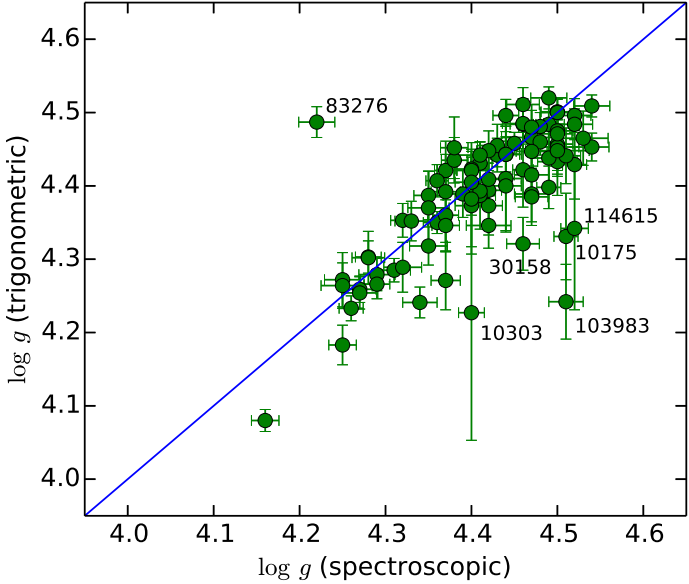}
\caption{Spectroscopic versus trigonometric surface gravity. The solid line corresponds to the 1:1 relation. Outlier stars are labeled with their HIP numbers.}
\label{f:logg}
\end{figure}

On average, the trigonometric $\logg$ error of our sample stars is 0.035. The mean $\logg$ difference ($\Delta(\logg)$, trigonometric minus spectroscopic) for our sample stars is $-0.02\pm0.07$. Given the formal errors in spectroscopic (0.019) and trigonometric (0.035) $\logg$ values, we would expect the 1\,$\sigma$ scatter of the $\logg$ differences to be lower (0.039). Figure~\ref{f:logg} shows a comparison of our spectroscopic and trigonometric $\logg$ values. There are a few outliers worth investigating. Six stars have $\Delta(\logg)$ greater than 2\,$\sigma$, where $\sigma$ is the standard deviation of the full sample's $\Delta(\logg)$ distribution. We discuss them in turn.

HIP\,10175 and HIP\,30158 have their $V$ magnitudes taken from the GCDP. These magnitudes are flagged as AB, which represents blended photometry, implying that their $V$ magnitudes are contaminated by a cooler nearby companion (no other star in our sample has the same AB flag in the GCDP). HIP\,10303 is in a wide binary system and its parallax has a very large error despite being a nearby system. The trigonometric $\logg$ of this star has an error of nearly 0.2. HIP\,114615 is the most distant and faintest star in our sample, which contributes to making the quality of both its spectroscopic and trigonometric $\logg$ values significantly below average. HIP\,83276 has been identified as a single-lined spectroscopic binary by \cite{duquennoy88}, who suggests a companion cooler than a K4-type dwarf. These authors derived a photometric parallax of about 31\,mas, which is well below the {\it Hipparcos} value ($36.5\pm1.4$\,mas), but will lower the trigonometric $\logg$ by only 0.1. HIP\,103983 is also a binary. \cite{tokovinin13} have identified a sub-arcsec companion which is affecting either the primary star's visual magnitude or the system's {\it Hipparcos} parallax (or both), leading to an incorrect trigonometric $\logg$. To confirm the latter, we estimated the age of the system using the {\it Hipparcos} $\logg$ instead of the spectroscopic value as was done in Section~\ref{s:massandage}. This results in $\sim9$\,Gyr; HIP\,103983's $\log R'_\mathrm{HK}=-4.84$ is too high for that age, but fully consistent with the solar $\log R'_\mathrm{HK}$ evolution for its ``spectroscopic'' age of about 2\,Gyr, as shown in Figure~\ref{f:age_rhk}.

\begin{figure}
\centering
\includegraphics[bb=115 240 500 565,width=9.1cm]{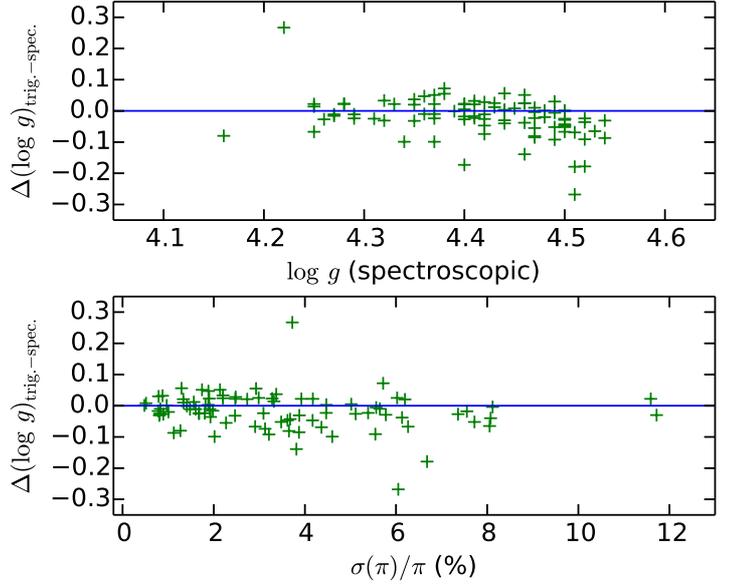}
\caption{Trigonometric minus spectroscopic surface gravity difference as a function of spectroscopic $\logg$ (top panel) and relative parallax error (bottom panel). The solid line is at zero.}
\label{f:dlogg}
\end{figure}

Excluding the six stars discussed above, $\Delta(\logg)$ reduces to $-0.01\pm0.04$. This difference is in principle fully explained by the formal errors. However, we note that $\Delta(\logg)$ exhibits minor trends with $\logg$ and in particular the relative error in the stellar parallax, as shown in Figure~\ref{f:dlogg}. No significant correlations were found for any of the other stellar parameters. The bottom panel of Figure~\ref{f:dlogg} shows that there is excellent agreement for $\delta(\pi)/\pi$ errors below 3\,\%, where $\pi$ is the {\it Hipparcos} trigonometric parallax. Indeed, the mean $\Delta(\logg)$ for those stars is $-0.004\pm0.037$ while that for the rest of our objects is $-0.026\pm0.040$ (excluding the 6 stars discussed in the previous paragraph).

Thus, we conclude that the higher uncertainty of the more distant stars in our sample leads to a small systematic difference between spectroscopic and trigonometric $\logg$ values. On the other hand, if we restrict the comparison to only the most precise trigonometric $\logg$ determinations, the agreement with our spectroscopic $\logg$ values is excellent.

\section{Conclusions}

We presented here the largest sample of solar twins (88 stars) analyzed homogeneously using high resolution, high signal-to-noise ratio spectra. Precise stellar parameters ($\teff$, $\logg$, [Fe/H], $\vt$) were obtained from a differential spectroscopic analysis relative to the Sun, which was observed using reflected light of asteroids, employing the same instrumentation and setup. We also measured stellar activity from the \ion{Ca}{ii} H and K lines.

Our stellar parameters have been validated using effective temperatures from the infrared flux method and from fits of $\ha$ line profiles, and with surface gravities determined using {\it Hipparcos} parallaxes. There is an excellent agreement with the independent determinations after their less precise cases are excluded from the comparisons, suggesting that systematic errors are negligible and that we can achieve the highest precision using the differential spectroscopic equilibrium, with effective temperatures determined to better than about 10\,K, $\logg$ with a precision of about 0.02\,dex, and [Fe/H] to better than about 0.01\,dex.

The precise atmospheric parameters ($\teff$, $\logg$, [Fe/H]) were used to determine isochrone masses and ages, taking the error bars in the determination of the stellar parameters into account. The masses are within about 5\,\% of the solar mass, and there is a range in ages of about 0.5--10\,Gyr. Our sample is ideal to test different aspects of the main-sequence evolution of the Sun, such as the evolution of surface lithium abundance with age \citep{baumann10,monroe13,melendez14:hip114328} or the decay of stellar activity.

Although the goal of this work was to present our ``Solar Twin Planet Search'' project and to provide the stellar parameters for our sample stars to use in future publications, two important scientific results were obtained while preparing this ``input catalog'':

\begin{itemize}

\item i) The formal errors in stellar parameters derived from strict differential analysis are excellent indicators of the actual uncertainty of those measurements. In other words, systematic errors in the derivation of fundamental atmospheric parameters using only the iron lines are negligible when studying solar twin stars. This had been assumed in our previous work (and similar works by other groups), but it has only now been demonstrated.

\item ii) A very tight $\log R'_\mathrm{HK}$ versus age relation is found for our sample of solar twin stars thanks to the high precision of our relative stellar ages. This trend can be employed to quantitatively constrain evolutionary models of stellar activity and rotation. The fact that the solar cycle fits this trend and its dispersion very well shows that the sample size is appropriate to take variations in the $\log R'_\mathrm{HK}$ index due to the stars' activity cycles into account, although that will be further improved once our multi-epoch HARPS data are analyzed in a similar way.

\end{itemize}

The stellar parameters ($\teff$, $\logg$, [Fe/H], $\vt$) and fundamental properties (mass, age, stellar activity) determined here will be employed in a subsequent series of papers aiming to obtain at high precision the detailed chemical composition of solar twins and to characterize the stars' planetary systems from our dedicated solar twin planet search. Furthermore, our sample will be also useful for other applications related to stellar astrophysics, such as constraining non-standard stellar models and studying the chemical evolution of our Galaxy.

\begin{acknowledgements}
I.R.\ acknowledges past support from NASA's Sagan Fellowship Program to conduct the MIKE/Magellan observations presented in this paper. J.M.\ would like to acknowledge support from FAPESP (2012/24392-2) and CNPq (Bolsa de Produtividade). J.B.\ and M.B.\ acknowledge support for this work from the National Science Foundation (NSF, grant number AST-1313119). M.B.\ acknowledges support from the NSF in the form of a Graduate Research Fellowship (grant number DGE-1144082). L.S.\ acknowledges support from FAPESP (2013/25008-4).
\end{acknowledgements}

\bibliographystyle{aa}

\onecolumn

{\small
\begin{longtable}{rrcccrc}
\caption{Sample of solar twin stars} \\ \hline\hline
HIP & HD & $V$ & UT of observation & JD & $RV$ & $\log R'_\mathrm{HK}$ \\
& & (mag) & (Y-M-D / H:M:S) & (days) & (\kms) & \\ \hline
\endfirsthead
\caption{continued.} \\ \hline\hline
HIP & HD & $V$ & UT of observation & RJD & $RV$ & $\log R'_\mathrm{HK}$ \\
& & (mag) & (Y-M-D / H:M:S) & (days) & (\kms) & \\ \hline
\endhead
\hline
\endfoot
1954   &   2071  & 7.27 & 2011-01-03 / 02:54:00 & 2455564.62303 & $  6.65\pm0.22$ & $-4.931$ \\
3203   &   3821A & 7.02 & 2011-06-24 / 06:32:18 & 2455736.77700 & $ 11.99\pm0.55$ & $-4.508$ \\
4909   &   6204  & 8.51 & 2011-06-24 / 07:03:30 & 2455736.79732 & $ -3.21\pm0.46$ & $-4.522$ \\
5301   &   6718  & 8.44 & 2011-09-09 / 01:39:13 & 2455813.57328 & $ 34.38\pm0.24$ & $-4.962$ \\
6407   &   8291  & 8.62 & 2011-01-04 / 01:05:54 & 2455565.54735 & $  6.49\pm0.29$ & $-4.740$ \\
7585   &   9986  & 6.77 & 2011-09-09 / 02:56:32 & 2455813.62749 & $-20.90\pm0.31$ & $-4.895$ \\
8507   &  11195  & 8.90 & 2011-09-09 / 03:29:09 & 2455813.65023 & $ -9.58\pm0.75$ & $-4.908$ \\
9349   &  12264  & 7.99 & 2011-09-09 / 04:01:34 & 2455813.67186 & $ 19.33\pm0.64$ & $-4.559$ \\
10175  &  13357  & 7.65 & 2011-09-09 / 04:19:53 & 2455813.68257 & $ 25.18\pm0.66$ & $-4.796$ \\
10303  &  13612B & 7.56 & 2011-06-23 / 08:10:00 & 2455735.84358 & $ -5.62\pm0.58$ & $-4.959$ \\
11915  &  16008  & 8.62 & 2011-09-09 / 04:43:03 & 2455813.70187 & $ 14.36\pm0.76$ & $-4.880$ \\
14501  &  19467  & 6.97 & 2011-09-09 / 05:16:19 & 2455813.72526 & $  7.68\pm0.95$ & $-4.999$ \\
14614  &  19518  & 7.84 & 2011-06-23 / 08:40:44 & 2455735.86506 & $-27.44\pm0.68$ & $-4.945$ \\
14623  &  19632  & 7.29 & 2011-06-23 / 09:43:40 & 2455735.90759 & $ 40.98\pm0.76$ & $-4.431$ \\
15527  &  20782  & 7.38 & 2011-01-03 / 00:08:56 & 2455564.50326 & $ 40.39\pm0.83$ & $-4.943$ \\
18844  &  25874  & 6.73 & 2011-01-02 / 01:02:01 & 2455563.54134 & $ 34.84\pm0.88$ & $-5.000$ \\
19911  &  26990  & 7.51 & 2011-01-02 / 01:31:28 & 2455563.56230 & $-14.73\pm0.80$ & $-4.620$ \\
21079  &  28904  & 8.26 & 2011-01-02 / 01:54:42 & 2455563.57813 & $ -6.05\pm0.95$ & $-4.498$ \\
22263  &  30495  & 5.50 & 2011-01-01 / 00:54:52 & 2455562.53961 & $ 22.23\pm0.90$ & $-4.532$ \\
22395  &  30774  & 7.88 & 2011-01-01 / 01:42:31 & 2455562.57362 & $ 17.61\pm0.27$ & $-4.452$ \\
25670  &  36152  & 8.28 & 2011-09-10 / 07:32:20 & 2455814.81687 & $ -9.91\pm0.15$ & $-4.896$ \\
28066  &  39881  & 6.60 & 2011-01-04 / 02:20:33 & 2455565.59924 & $  1.09\pm0.88$ & $-5.004$ \\
29432  &  42618  & 6.87 & 2011-01-01 / 02:14:31 & 2455562.59521 & $-53.67\pm0.21$ & $-4.948$ \\
29525  &  42807  & 6.44 & 2011-01-01 / 02:41:19 & 2455562.61198 & $  6.08\pm0.23$ & $-4.445$ \\
30037  &  45021  & 9.16 & 2011-01-02 / 02:31:47 & 2455563.60329 & $ 49.79\pm0.69$ & $-4.929$ \\
30158  &  44665  & 8.37 & 2011-01-01 / 03:06:38 & 2455562.63401 & $ 44.57\pm0.67$ & $-5.015$ \\
30344  &  44821  & 7.37 & 2011-01-01 / 03:42:10 & 2455562.65576 & $ 14.66\pm0.69$ & $-4.470$ \\
30476  &  45289  & 6.65 & 2011-01-01 / 03:30:26 & 2455562.64984 & $ 57.11\pm0.93$ & $-5.012$ \\
30502  &  45346  & 8.67 & 2011-01-04 / 03:37:12 & 2455565.65346 & $ 45.53\pm0.52$ & $-4.971$ \\
33094  &  50806  & 6.03 & 2011-09-09 / 07:24:42 & 2455813.80852 & $ 72.75\pm0.32$ & $-5.058$ \\
34511  &  54351  & 8.00 & 2011-01-01 / 04:15:07 & 2455562.68265 & $ 42.91\pm0.69$ & $-4.953$ \\
36512  &  59711  & 7.73 & 2011-01-02 / 04:31:37 & 2455563.69400 & $  8.45\pm0.82$ & $-4.946$ \\
36515  &  59967  & 6.64 & 2011-01-02 / 04:52:54 & 2455563.70888 & $  9.74\pm0.39$ & $-4.396$ \\
38072  &  63487  & 9.22 & 2012-02-23 / 01:25:37 & 2455980.55951 & $ 58.54\pm0.68$ & $-4.472$ \\
40133  &  68168  & 7.36 & 2012-02-23 / 00:22:44 & 2455980.51704 & $  9.40\pm0.38$ & $-4.979$ \\
41317  &  71334  & 7.81 & 2011-01-03 / 04:42:49 & 2455564.70020 & $ 18.26\pm1.07$ & $-4.970$ \\
42333  &  73350  & 6.75 & 2011-01-01 / 04:35:47 & 2455562.69382 & $ 35.30\pm0.20$ & $-4.533$ \\
43297  &  75302  & 7.46 & 2012-04-29 / 22:52:02 & 2456047.45169 & $ 10.50\pm0.52$ & $-4.648$ \\
44713  &  78429  & 7.31 & 2011-06-24 / 10:12:56 & 2455736.92526 & $ 65.39\pm0.78$ & $-4.951$ \\
44935  &  78534  & 8.74 & 2011-01-02 / 05:23:06 & 2455563.72791 & $-37.99\pm0.79$ & $-4.993$ \\
44997  &  78660  & 8.35 & 2012-02-23 / 02:34:17 & 2455980.61100 & $-10.68\pm0.71$ & $-4.966$ \\
49756  &  88072  & 7.54 & 2011-01-01 / 05:23:01 & 2455562.72863 & $-17.89\pm0.48$ & $-4.962$ \\
54102  &  96116  & 8.65 & 2011-01-02 / 05:41:19 & 2455563.73977 & $ 30.93\pm0.61$ & $-4.719$ \\
54287  &  96423  & 7.23 & 2012-02-23 / 03:23:58 & 2455980.64507 & $ 55.75\pm0.97$ & $-5.008$ \\
54582  &  97037  & 6.81 & 2011-01-04 / 07:20:48 & 2455565.81150 & $-15.12\pm0.98$ & $-5.001$ \\
55409  &  98649  & 8.00 & 2011-01-01 / 06:31:00 & 2455562.77452 & $  3.61\pm0.59$ & $-4.973$ \\
62039  & 110537  & 7.84 & 2011-01-02 / 08:32:30 & 2455563.86027 & $ 36.38\pm0.73$ & $-5.028$ \\
64150  & 114174  & 6.82 & 2011-01-03 / 06:34:45 & 2455564.77884 & $ 24.08\pm0.74$ & $-4.991$ \\
64673  & 115031  & 8.34 & 2011-01-01 / 06:53:16 & 2455562.78856 & $ 69.11\pm0.68$ & $-4.972$ \\
64713  & 115169  & 9.26 & 2011-01-01 / 07:36:34 & 2455562.82163 & $ 20.27\pm0.83$ & $-4.972$ \\
65708  & 117126  & 7.43 & 2012-04-29 / 23:46:52 & 2456047.49144 & $-14.22\pm0.76$ & $-5.021$ \\
67620  & 120690  & 6.44 & 2011-09-09 / 05:32:36 & 2455813.73608 & $  1.80\pm0.59$ & $-4.797$ \\
68468  & 122194  & 9.39 & 2011-01-04 / 07:59:38 & 2455565.83670 & $  0.38\pm0.77$ & $-5.022$ \\
69645  & 124523  & 9.41 & 2011-01-03 / 01:30:00 & 2455564.56251 & $ -3.68\pm0.76$ & $-4.975$ \\
72043  & 129814  & 7.53 & 2012-04-30 / 00:37:39 & 2456047.52927 & $  6.01\pm0.64$ & $-4.928$ \\
73241  & 131923  & 6.35 & 2011-01-01 / 08:23:40 & 2455562.84979 & $ 32.90\pm0.65$ & $-5.001$ \\
73815  & 133600  & 8.17 & 2011-01-01 / 08:44:32 & 2455562.86608 & $  3.93\pm0.61$ & $-4.990$ \\
74389  & 134664  & 7.77 & 2012-04-30 / 01:42:53 & 2456047.57552 & $  7.61\pm0.77$ & $-4.935$ \\
74432  & 135101  & 6.68 & 2011-06-23 / 23:28:57 & 2455736.47939 & $-39.77\pm0.71$ & $-5.042$ \\
76114  & 138573  & 7.23 & 2011-09-09 / 06:01:22 & 2455813.75575 & $-36.47\pm0.89$ & $-4.984$ \\
77052  & 140538  & 5.87 & 2012-04-30 / 02:16:12 & 2456047.59969 & $ 17.93\pm0.89$ & $-4.834$ \\
77883  & 142331  & 8.76 & 2012-04-30 / 02:24:26 & 2456047.60455 & $-70.94\pm0.66$ & $-4.990$ \\
79578  & 145825  & 6.53 & 2012-05-01 / 00:14:06 & 2456048.51514 & $-21.20\pm0.34$ & $-4.732$ \\
79672  & 146233  & 5.51 & 2012-04-30 / 03:29:46 & 2456047.65110 & $ 10.98\pm0.85$ & $-4.983$ \\
79715  & 145927  & 8.36 & 2012-04-30 / 04:59:59 & 2456047.71317 & $ 14.12\pm0.69$ & $-4.980$ \\
81746  & 150248  & 7.03 & 2011-06-24 / 01:09:18 & 2455736.55225 & $ 67.26\pm0.37$ & $-4.962$ \\
83276  & 153631  & 7.13 & 2012-04-30 / 05:58:46 & 2456047.75450 & $ 88.46\pm0.34$ & $-4.990$ \\
85042  & 157347  & 6.29 & 2011-06-24 / 01:27:22 & 2455736.56517 & $-36.46\pm0.57$ & $-5.014$ \\
87769  & 163441  & 8.44 & 2012-04-30 / 04:20:30 & 2456047.68563 & $ 12.05\pm0.70$ & $-4.953$ \\
89650  & 167060  & 8.94 & 2011-01-03 / 02:04:59 & 2455564.58823 & $ 15.01\pm0.61$ & $-4.997$ \\
95962  & 183658  & 7.28 & 2012-04-30 / 04:40:53 & 2456047.70009 & $ 57.66\pm0.62$ & $-4.985$ \\
96160  & 183579  & 8.69 & 2011-06-24 / 02:18:52 & 2455736.60066 & $-15.86\pm0.38$ & $-4.855$ \\
101905 & 196390  & 7.33 & 2012-04-30 / 07:22:32 & 2456047.81274 & $ 27.12\pm0.69$ & $-4.731$ \\
102040 & 197076A & 6.44 & 2011-09-08 / 23:19:07 & 2455813.47075 & $-36.21\pm0.65$ & $-4.927$ \\
102152 & 197027  & 9.21 & 2012-02-23 / 09:26:19 & 2455980.89334 & $-44.02\pm0.16$ & $-4.989$ \\
103983 & 200565  & 8.45 & 2012-04-30 / 07:50:57 & 2456047.83226 & $ -2.71\pm0.54$ & $-4.843$ \\
104045 & 200633  & 8.41 & 2012-05-01 / 03:06:25 & 2456048.63470 & $ 45.00\pm0.25$ & $-4.969$ \\
105184 & 202628  & 6.74 & 2011-06-24 / 03:15:19 & 2455736.64065 & $ 11.79\pm0.46$ & $-4.720$ \\
108158 & 207700  & 7.42 & 2011-09-08 / 23:34:42 & 2455813.48209 & $  6.37\pm0.39$ & $-4.974$ \\
108468 & 208704  & 7.21 & 2011-09-09 / 00:21:41 & 2455813.51498 & $  3.32\pm0.61$ & $-4.975$ \\
108996 & 209562  & 8.88 & 2011-09-09 / 00:38:13 & 2455813.52649 & $ 12.26\pm0.57$ & $-4.466$ \\
109110 & 209779  & 7.57 & 2011-06-24 / 03:33:29 & 2455736.65361 & $-10.79\pm0.38$ & $-4.490$ \\
109821 & 210918  & 6.23 & 2011-09-10 / 05:47:08 & 2455814.74547 & $-19.24\pm0.58$ & $-5.004$ \\
114615 & 219057  & 9.59 & 2012-04-30 / 06:25:21 & 2456047.77076 & $  0.55\pm0.61$ & $-4.838$ \\
115577 & 220507  & 7.60 & 2012-04-30 / 08:03:52 & 2456047.83881 & $ 23.85\pm0.85$ & $-5.020$ \\
116906 & 222582  & 7.68 & 2011-09-10 / 06:51:23 & 2455814.79000 & $ 12.15\pm0.21$ & $-4.978$ \\
117367 & 223238  & 7.71 & 2011-06-23 / 06:23:10 & 2455735.77134 & $-15.37\pm0.26$ & $-5.013$ \\
118115 & 224383  & 7.85 & 2011-09-09 / 00:57:17 & 2455813.54221 & $-31.23\pm0.16$ & $-4.995$ \\

\label{t:sample}
\end{longtable}
}

\newpage

{\small
\begin{longtable}{ccccr}
\caption{Iron line list} \\ \hline\hline
Wavelength & Species & $\chi$ & $\log(gf)$ & $EW_\odot$ \\
(\AA) & & (eV) & & (m\AA) \\ \hline
\endfirsthead
\caption{continued.} \\ \hline\hline
Wavelength & Species & $\chi$ & $\log(gf)$ & $EW_\odot$ \\
(\AA) & & (eV) & & (m\AA) \\ \hline
\endhead
\hline
\endfoot
4389.25 & 26.0 & 0.05 & $-4.58$ & $73.2\pm0.1$ \\
4445.47 & 26.0 & 0.09 & $-5.44$ & $40.4\pm0.2$ \\
4602.00 & 26.0 & 1.61 & $-3.15$ & $72.3\pm0.5$ \\
4690.14 & 26.0 & 3.69 & $-1.61$ & $59.5\pm0.3$ \\
4788.76 & 26.0 & 3.24 & $-1.73$ & $67.7\pm0.4$ \\
4799.41 & 26.0 & 3.64 & $-2.13$ & $36.0\pm0.6$ \\
4808.15 & 26.0 & 3.25 & $-2.69$ & $27.6\pm0.3$ \\
4950.10 & 26.0 & 3.42 & $-1.56$ & $74.6\pm0.3$ \\
4994.13 & 26.0 & 0.92 & $-3.08$ & $102.0\pm0.4$ \\
5141.74 & 26.0 & 2.42 & $-2.23$ & $90.6\pm0.9$ \\
5198.71 & 26.0 & 2.22 & $-2.14$ & $99.2\pm0.4$ \\
5225.53 & 26.0 & 0.11 & $-4.79$ & $74.9\pm0.4$ \\
5242.49 & 26.0 & 3.63 & $-0.99$ & $87.7\pm1.2$ \\
5247.05 & 26.0 & 0.09 & $-4.96$ & $67.7\pm0.2$ \\
5250.21 & 26.0 & 0.12 & $-4.94$ & $66.8\pm0.2$ \\
5295.31 & 26.0 & 4.42 & $-1.59$ & $30.3\pm0.5$ \\
5322.04 & 26.0 & 2.28 & $-2.89$ & $62.3\pm0.2$ \\
5373.71 & 26.0 & 4.47 & $-0.74$ & $63.6\pm0.3$ \\
5379.57 & 26.0 & 3.69 & $-1.51$ & $62.4\pm0.1$ \\
5386.33 & 26.0 & 4.15 & $-1.67$ & $33.0\pm0.1$ \\
5441.34 & 26.0 & 4.31 & $-1.63$ & $32.3\pm0.7$ \\
5466.40 & 26.0 & 4.37 & $-0.57$ & $78.8\pm0.7$ \\
5466.99 & 26.0 & 3.57 & $-2.23$ & $34.3\pm0.3$ \\
5491.83 & 26.0 & 4.19 & $-2.19$ & $14.0\pm0.2$ \\
5554.89 & 26.0 & 4.55 & $-0.36$ & $98.8\pm0.6$ \\
5560.21 & 26.0 & 4.43 & $-1.09$ & $53.4\pm0.6$ \\
5618.63 & 26.0 & 4.21 & $-1.27$ & $51.3\pm0.3$ \\
5638.26 & 26.0 & 4.22 & $-0.77$ & $78.5\pm0.4$ \\
5651.47 & 26.0 & 4.47 & $-1.75$ & $19.2\pm0.2$ \\
5679.02 & 26.0 & 4.65 & $-0.75$ & $60.0\pm0.3$ \\
5701.54 & 26.0 & 2.56 & $-2.16$ & $86.4\pm0.1$ \\
5705.46 & 26.0 & 4.30 & $-1.36$ & $39.4\pm0.5$ \\
5731.76 & 26.0 & 4.26 & $-1.20$ & $58.7\pm0.1$ \\
5775.08 & 26.0 & 4.22 & $-1.30$ & $60.3\pm0.4$ \\
5778.45 & 26.0 & 2.59 & $-3.44$ & $22.9\pm0.4$ \\
5784.66 & 26.0 & 3.40 & $-2.53$ & $27.7\pm0.3$ \\
5793.91 & 26.0 & 4.22 & $-1.62$ & $35.0\pm0.4$ \\
5806.73 & 26.0 & 4.61 & $-0.95$ & $55.4\pm0.5$ \\
5852.22 & 26.0 & 4.55 & $-1.23$ & $41.3\pm0.4$ \\
5855.08 & 26.0 & 4.61 & $-1.48$ & $23.3\pm0.4$ \\
5930.18 & 26.0 & 4.65 & $-0.17$ & $89.6\pm0.1$ \\
5934.65 & 26.0 & 3.93 & $-1.07$ & $77.6\pm0.7$ \\
5956.69 & 26.0 & 0.86 & $-4.55$ & $52.6\pm1.2$ \\
5987.07 & 26.0 & 4.80 & $-0.21$ & $70.9\pm1.0$ \\
6003.01 & 26.0 & 3.88 & $-1.06$ & $85.7\pm1.0$ \\
6005.54 & 26.0 & 2.59 & $-3.43$ & $22.7\pm0.3$ \\
6027.05 & 26.0 & 4.08 & $-1.09$ & $65.1\pm0.4$ \\
6056.00 & 26.0 & 4.73 & $-0.40$ & $74.6\pm0.3$ \\
6065.48 & 26.0 & 2.61 & $-1.53$ & $119.5\pm0.4$ \\
6079.01 & 26.0 & 4.65 & $-1.02$ & $47.5\pm0.1$ \\
6082.71 & 26.0 & 2.22 & $-3.57$ & $36.1\pm0.1$ \\
6093.64 & 26.0 & 4.61 & $-1.30$ & $31.5\pm0.2$ \\
6096.67 & 26.0 & 3.98 & $-1.81$ & $38.8\pm0.2$ \\
6151.62 & 26.0 & 2.18 & $-3.28$ & $51.1\pm0.3$ \\
6165.36 & 26.0 & 4.14 & $-1.46$ & $45.7\pm0.2$ \\
6173.34 & 26.0 & 2.22 & $-2.88$ & $69.4\pm0.3$ \\
6187.99 & 26.0 & 3.94 & $-1.62$ & $48.9\pm0.5$ \\
6200.31 & 26.0 & 2.61 & $-2.42$ & $74.4\pm0.3$ \\
6213.43 & 26.0 & 2.22 & $-2.52$ & $83.9\pm0.2$ \\
6219.28 & 26.0 & 2.20 & $-2.43$ & $90.9\pm0.7$ \\
6226.74 & 26.0 & 3.88 & $-2.10$ & $30.2\pm0.4$ \\
6232.64 & 26.0 & 3.65 & $-1.22$ & $85.9\pm0.6$ \\
6240.65 & 26.0 & 2.22 & $-3.29$ & $50.1\pm0.5$ \\
6265.13 & 26.0 & 2.18 & $-2.55$ & $87.5\pm0.5$ \\
6271.28 & 26.0 & 3.33 & $-2.70$ & $25.4\pm0.4$ \\
6322.69 & 26.0 & 2.59 & $-2.43$ & $77.5\pm0.1$ \\
6380.74 & 26.0 & 4.19 & $-1.32$ & $53.2\pm0.3$ \\
6392.54 & 26.0 & 2.28 & $-4.03$ & $17.8\pm0.4$ \\
6430.85 & 26.0 & 2.18 & $-2.01$ & $113.1\pm0.4$ \\
6498.94 & 26.0 & 0.96 & $-4.70$ & $46.4\pm0.4$ \\
6593.87 & 26.0 & 2.43 & $-2.39$ & $85.7\pm0.9$ \\
6597.56 & 26.0 & 4.80 & $-0.97$ & $45.3\pm0.4$ \\
6625.02 & 26.0 & 1.01 & $-5.34$ & $15.9\pm0.2$ \\
6703.57 & 26.0 & 2.76 & $-3.02$ & $38.2\pm0.3$ \\
6705.10 & 26.0 & 4.61 & $-0.98$ & $47.7\pm1.0$ \\
6710.32 & 26.0 & 1.49 & $-4.88$ & $16.1\pm0.2$ \\
6713.75 & 26.0 & 4.80 & $-1.40$ & $21.5\pm0.4$ \\
6725.36 & 26.0 & 4.10 & $-2.19$ & $18.1\pm0.4$ \\
6726.67 & 26.0 & 4.61 & $-1.03$ & $47.9\pm0.2$ \\
6733.15 & 26.0 & 4.64 & $-1.47$ & $27.6\pm0.3$ \\
6739.52 & 26.0 & 1.56 & $-4.79$ & $12.5\pm0.6$ \\
6750.15 & 26.0 & 2.42 & $-2.62$ & $75.1\pm0.2$ \\
6793.26 & 26.0 & 4.08 & $-2.33$ & $13.7\pm0.3$ \\
6806.85 & 26.0 & 2.73 & $-3.11$ & $35.7\pm0.1$ \\
6810.26 & 26.0 & 4.61 & $-0.99$ & $51.7\pm0.9$ \\
6837.01 & 26.0 & 4.59 & $-1.69$ & $17.2\pm0.6$ \\
6839.83 & 26.0 & 2.56 & $-3.35$ & $30.6\pm0.1$ \\
6843.66 & 26.0 & 4.55 & $-0.83$ & $62.8\pm0.9$ \\
6858.15 & 26.0 & 4.61 & $-0.94$ & $52.7\pm0.7$ \\
7583.79 & 26.0 & 3.02 & $-1.88$ & $84.9\pm0.4$ \\
7723.21 & 26.0 & 2.28 & $-3.62$ & $44.6\pm0.7$ \\
4491.40 & 26.1 & 2.86 & $-2.66$ & $79.2\pm0.6$ \\
4508.29 & 26.1 & 2.86 & $-2.52$ & $87.8\pm0.7$ \\
4576.33 & 26.1 & 2.84 & $-2.95$ & $64.9\pm0.2$ \\
4620.51 & 26.1 & 2.83 & $-3.21$ & $54.7\pm0.4$ \\
4993.34 & 26.1 & 2.81 & $-3.73$ & $36.8\pm0.3$ \\
5197.58 & 26.1 & 3.23 & $-2.22$ & $83.4\pm0.2$ \\
5234.62 & 26.1 & 3.22 & $-2.18$ & $84.5\pm0.5$ \\
5264.80 & 26.1 & 3.23 & $-3.13$ & $45.5\pm0.2$ \\
5325.55 & 26.1 & 3.22 & $-3.25$ & $41.3\pm0.2$ \\
5414.07 & 26.1 & 3.22 & $-3.58$ & $28.0\pm0.9$ \\
5425.26 & 26.1 & 3.20 & $-3.22$ & $41.5\pm0.1$ \\
6084.09 & 26.1 & 3.20 & $-3.83$ & $21.1\pm0.4$ \\
6149.24 & 26.1 & 3.89 & $-2.75$ & $36.2\pm0.3$ \\
6247.55 & 26.1 & 3.89 & $-2.38$ & $53.5\pm0.4$ \\
6369.46 & 26.1 & 2.89 & $-4.11$ & $19.7\pm0.4$ \\
6416.92 & 26.1 & 3.89 & $-2.75$ & $40.7\pm0.4$ \\
6432.68 & 26.1 & 2.89 & $-3.57$ & $42.4\pm0.8$ \\
6456.38 & 26.1 & 3.90 & $-2.05$ & $65.3\pm0.1$ \\
7515.83 & 26.1 & 3.90 & $-3.39$ & $14.4\pm0.5$ \\

\label{t:linelist}
\end{longtable}
}

\newpage

{\small
\begin{landscape}
\begin{longtable}{rrrrrrrrrrrrrrr}
\caption{Stellar parameters} \\ \hline\hline
HIP & HD & $\teff$ & $\sigma(\teff)$ & $\logg$ & $\sigma(\logg)$ & $\feh$ & $\sigma(\feh)$ & $\vt$ & $\sigma(\vt)$ & $M$ & $\sigma(M)$ & $\tau$ & $\tau_{-\sigma}$ & $\tau_{+\sigma}$ \\ 
& & (K) & (K) & ([cgs]) & ([cgs]) & (dex) & (dex) & (\kms) & (\kms) & ($M_\odot$) & ($M_\odot$) & (Gyr) & (Gyr) & (Gyr) \\ \hline
\endfirsthead
\caption{continued.} \\ \hline\hline
HIP & HD & $\teff$ & $\sigma(\teff)$ & $\logg$ & $\sigma(\logg)$ & $\feh$ & $\sigma(\feh)$ & $\vt$ & $\sigma(\vt)$ & $M$ & $\sigma(M)$ & $\tau$ & $\tau_{-\sigma}$ & $\tau_{+\sigma}$ \\ 
& & (K) & (K) & ([cgs]) & ([cgs]) & (dex) & (dex) & (\kms) & (\kms) & ($M_\odot$) & ($M_\odot$) & (Gyr) & (Gyr) & (Gyr) \\ \hline
\endhead
\hline
\endfoot
1954 & 2071 & 5717 & 5 & 4.46 & 0.02 & $-0.068$ & 0.006 & 0.96 & 0.02 & 0.974 & 0.007 & 4.7 & 3.9 & 5.7 \\
3203 & 3821A & 5850 & 10 & 4.52 & 0.02 & $-0.087$ & 0.008 & 1.16 & 0.02 & 1.034 & 0.008 & 0.8 & 0.3 & 1.6 \\
4909 & 6204 & 5854 & 10 & 4.50 & 0.02 & $0.028$ & 0.008 & 1.12 & 0.02 & 1.052 & 0.007 & 1.2 & 0.0 & 2.0 \\
5301 & 6718 & 5728 & 5 & 4.42 & 0.02 & $-0.064$ & 0.004 & 0.97 & 0.01 & 0.967 & 0.005 & 6.4 & 5.7 & 7.0 \\
6407 & 8291 & 5764 & 8 & 4.52 & 0.01 & $-0.068$ & 0.007 & 0.97 & 0.02 & 1.005 & 0.005 & 0.9 & 0.6 & 1.9 \\
7585 & 9986 & 5831 & 5 & 4.43 & 0.01 & $0.095$ & 0.005 & 1.02 & 0.01 & 1.060 & 0.002 & 3.3 & 2.7 & 3.7 \\
8507 & 11195 & 5725 & 6 & 4.49 & 0.02 & $-0.096$ & 0.006 & 0.99 & 0.02 & 0.978 & 0.007 & 3.6 & 2.7 & 4.5 \\
9349 & 12264 & 5810 & 8 & 4.50 & 0.02 & $0.009$ & 0.007 & 1.07 & 0.02 & 1.038 & 0.010 & 0.5 & 0.3 & 1.9 \\
10175 & 13357 & 5738 & 7 & 4.51 & 0.01 & $-0.007$ & 0.005 & 0.96 & 0.01 & 1.011 & 0.005 & 2.0 & 1.0 & 2.6 \\
10303 & 13612B & 5725 & 4 & 4.40 & 0.01 & $0.106$ & 0.004 & 0.98 & 0.01 & 1.029 & 0.003 & 5.3 & 4.6 & 5.7 \\
11915 & 16008 & 5760 & 4 & 4.46 & 0.01 & $-0.059$ & 0.004 & 0.97 & 0.01 & 0.993 & 0.005 & 4.0 & 3.4 & 4.6 \\
14501 & 19467 & 5728 & 7 & 4.29 & 0.02 & $-0.133$ & 0.005 & 1.03 & 0.01 & 0.951 & 0.004 & 9.9 & 9.5 & 10.3 \\
14614 & 19518 & 5784 & 9 & 4.42 & 0.03 & $-0.099$ & 0.008 & 1.03 & 0.02 & 0.976 & 0.008 & 6.1 & 4.8 & 6.9 \\
14623 & 19632 & 5769 & 13 & 4.52 & 0.02 & $0.106$ & 0.010 & 1.15 & 0.02 & 1.051 & 0.008 & 1.1 & 0.5 & 1.8 \\
15527 & 20782 & 5785 & 5 & 4.32 & 0.01 & $-0.051$ & 0.005 & 1.05 & 0.01 & 0.993 & 0.005 & 7.8 & 7.5 & 8.1 \\
18844 & 25874 & 5736 & 5 & 4.36 & 0.02 & $0.016$ & 0.004 & 0.99 & 0.01 & 0.989 & 0.003 & 7.3 & 6.9 & 7.7 \\
19911 & 26990 & 5764 & 12 & 4.47 & 0.04 & $-0.070$ & 0.011 & 1.02 & 0.03 & 0.987 & 0.012 & 4.0 & 2.4 & 5.5 \\
21079 & 28904 & 5846 & 11 & 4.50 & 0.03 & $-0.070$ & 0.008 & 1.09 & 0.02 & 1.026 & 0.009 & 0.8 & 0.4 & 2.4 \\
22263 & 30495 & 5840 & 8 & 4.50 & 0.02 & $0.030$ & 0.007 & 1.08 & 0.02 & 1.053 & 0.007 & 0.5 & 0.2 & 1.6 \\
22395 & 30774 & 5789 & 8 & 4.43 & 0.02 & $0.084$ & 0.008 & 1.11 & 0.02 & 1.033 & 0.006 & 4.0 & 3.1 & 4.7 \\
25670 & 36152 & 5771 & 5 & 4.44 & 0.02 & $0.057$ & 0.005 & 1.00 & 0.01 & 1.025 & 0.005 & 4.5 & 3.4 & 4.9 \\
28066 & 39881 & 5733 & 5 & 4.29 & 0.01 & $-0.128$ & 0.004 & 1.05 & 0.01 & 0.959 & 0.003 & 9.7 & 9.4 & 10.0 \\
29432 & 42618 & 5758 & 5 & 4.44 & 0.01 & $-0.096$ & 0.005 & 1.01 & 0.01 & 0.974 & 0.007 & 5.4 & 4.6 & 6.0 \\
29525 & 42807 & 5737 & 7 & 4.49 & 0.02 & $-0.022$ & 0.007 & 1.12 & 0.02 & 0.997 & 0.007 & 2.9 & 1.8 & 4.0 \\
30037 & 45021 & 5668 & 5 & 4.42 & 0.01 & $-0.011$ & 0.004 & 0.94 & 0.01 & 0.961 & 0.004 & 6.7 & 6.1 & 7.3 \\
30158 & 44665 & 5702 & 5 & 4.46 & 0.02 & $0.003$ & 0.006 & 0.94 & 0.02 & 0.988 & 0.008 & 4.5 & 3.4 & 5.4 \\
30344 & 44821 & 5750 & 9 & 4.50 & 0.02 & $0.063$ & 0.007 & 1.09 & 0.02 & 1.026 & 0.005 & 1.2 & 0.8 & 2.4 \\
30476 & 45289 & 5710 & 5 & 4.26 & 0.01 & $-0.022$ & 0.004 & 1.03 & 0.01 & 0.980 & 0.002 & 9.5 & 9.3 & 9.8 \\
30502 & 45346 & 5721 & 6 & 4.41 & 0.02 & $-0.076$ & 0.006 & 0.98 & 0.02 & 0.960 & 0.003 & 6.8 & 6.1 & 7.5 \\
33094 & 50806 & 5662 & 7 & 4.16 & 0.02 & $0.043$ & 0.005 & 1.13 & 0.01 & 1.005 & 0.007 & 9.9 & 9.6 & 10.2 \\
34511 & 54351 & 5819 & 6 & 4.47 & 0.02 & $-0.103$ & 0.006 & 1.03 & 0.02 & 1.005 & 0.007 & 3.2 & 2.5 & 4.1 \\
36512 & 59711 & 5737 & 4 & 4.41 & 0.01 & $-0.117$ & 0.004 & 0.99 & 0.01 & 0.957 & 0.004 & 6.8 & 6.4 & 7.4 \\
36515 & 59967 & 5847 & 12 & 4.54 & 0.02 & $-0.021$ & 0.009 & 1.17 & 0.02 & 1.046 & 0.006 & 0.6 & 0.0 & 1.1 \\
38072 & 63487 & 5849 & 8 & 4.49 & 0.02 & $0.058$ & 0.007 & 1.14 & 0.02 & 1.059 & 0.005 & 0.9 & 0.5 & 1.9 \\
40133 & 68168 & 5755 & 4 & 4.37 & 0.01 & $0.128$ & 0.004 & 1.01 & 0.01 & 1.050 & 0.003 & 5.1 & 4.8 & 5.4 \\
41317 & 71334 & 5700 & 5 & 4.38 & 0.01 & $-0.068$ & 0.004 & 0.96 & 0.01 & 0.958 & 0.004 & 8.0 & 7.6 & 8.5 \\
42333 & 73350 & 5848 & 8 & 4.50 & 0.02 & $0.138$ & 0.008 & 1.16 & 0.02 & 1.086 & 0.005 & 1.0 & 0.4 & 1.5 \\
43297 & 75302 & 5702 & 5 & 4.46 & 0.01 & $0.083$ & 0.006 & 0.99 & 0.02 & 1.011 & 0.004 & 3.5 & 3.0 & 4.4 \\
44713 & 78429 & 5768 & 6 & 4.28 & 0.01 & $0.088$ & 0.005 & 1.06 & 0.01 & 1.039 & 0.003 & 7.3 & 7.0 & 7.6 \\
44935 & 78534 & 5782 & 5 & 4.37 & 0.01 & $0.058$ & 0.005 & 1.04 & 0.01 & 1.020 & 0.001 & 6.1 & 5.6 & 6.5 \\
44997 & 78660 & 5731 & 5 & 4.47 & 0.02 & $-0.023$ & 0.005 & 0.95 & 0.01 & 0.995 & 0.008 & 3.7 & 2.8 & 4.7 \\
49756 & 88072 & 5795 & 4 & 4.42 & 0.01 & $0.043$ & 0.004 & 1.01 & 0.01 & 1.022 & 0.004 & 4.6 & 4.0 & 5.0 \\
54102 & 96116 & 5820 & 9 & 4.51 & 0.02 & $-0.014$ & 0.007 & 1.02 & 0.02 & 1.035 & 0.007 & 0.5 & 0.3 & 1.7 \\
54287 & 96423 & 5727 & 4 & 4.36 & 0.01 & $0.118$ & 0.004 & 1.01 & 0.01 & 1.035 & 0.005 & 6.0 & 5.6 & 6.3 \\
54582 & 97037 & 5875 & 7 & 4.27 & 0.02 & $-0.080$ & 0.005 & 1.17 & 0.01 & 1.032 & 0.006 & 7.1 & 6.8 & 7.4 \\
55409 & 98649 & 5700 & 6 & 4.40 & 0.02 & $-0.080$ & 0.006 & 0.94 & 0.02 & 0.950 & 0.001 & 7.9 & 7.1 & 8.4 \\
62039 & 110537 & 5753 & 6 & 4.35 & 0.02 & $0.088$ & 0.005 & 1.05 & 0.01 & 1.028 & 0.004 & 6.3 & 5.9 & 6.8 \\
64150 & 114174 & 5747 & 6 & 4.39 & 0.02 & $0.030$ & 0.007 & 1.00 & 0.02 & 0.991 & 0.004 & 6.4 & 5.7 & 7.0 \\
64673 & 115031 & 5918 & 8 & 4.35 & 0.02 & $-0.030$ & 0.007 & 1.21 & 0.02 & 1.048 & 0.007 & 5.4 & 4.7 & 5.7 \\
64713 & 115169 & 5767 & 8 & 4.46 & 0.02 & $-0.067$ & 0.007 & 1.00 & 0.02 & 0.989 & 0.009 & 4.0 & 3.0 & 5.2 \\
65708 & 117126 & 5755 & 6 & 4.25 & 0.02 & $-0.066$ & 0.006 & 1.09 & 0.01 & 0.987 & 0.005 & 9.2 & 8.9 & 9.5 \\
67620 & 120690 & 5670 & 9 & 4.41 & 0.03 & $-0.018$ & 0.009 & 1.01 & 0.03 & 0.954 & 0.006 & 7.5 & 6.2 & 8.3 \\
68468 & 122194 & 5845 & 6 & 4.37 & 0.02 & $0.054$ & 0.005 & 1.13 & 0.01 & 1.040 & 0.002 & 5.2 & 4.7 & 5.6 \\
69645 & 124523 & 5743 & 6 & 4.44 & 0.02 & $-0.045$ & 0.006 & 0.99 & 0.02 & 0.981 & 0.008 & 5.2 & 4.3 & 6.0 \\
72043 & 129814 & 5842 & 8 & 4.35 & 0.02 & $-0.034$ & 0.007 & 1.12 & 0.02 & 1.015 & 0.006 & 6.4 & 5.9 & 6.8 \\
73241 & 131923 & 5669 & 8 & 4.27 & 0.02 & $0.082$ & 0.007 & 1.01 & 0.02 & 0.991 & 0.004 & 9.3 & 8.9 & 9.6 \\
73815 & 133600 & 5788 & 6 & 4.37 & 0.02 & $0.004$ & 0.005 & 1.05 & 0.01 & 1.003 & 0.005 & 6.5 & 6.0 & 6.9 \\
74389 & 134664 & 5844 & 5 & 4.49 & 0.01 & $0.077$ & 0.004 & 1.07 & 0.01 & 1.065 & 0.005 & 0.9 & 0.7 & 1.7 \\
74432 & 135101 & 5684 & 8 & 4.25 & 0.02 & $0.037$ & 0.007 & 1.09 & 0.02 & 0.986 & 0.005 & 9.6 & 9.3 & 9.9 \\
76114 & 138573 & 5733 & 6 & 4.42 & 0.02 & $-0.037$ & 0.006 & 0.97 & 0.02 & 0.977 & 0.007 & 6.1 & 5.2 & 6.8 \\
77052 & 140538 & 5683 & 5 & 4.48 & 0.02 & $0.036$ & 0.006 & 0.96 & 0.02 & 0.993 & 0.007 & 3.2 & 2.4 & 4.3 \\
77883 & 142331 & 5690 & 6 & 4.40 & 0.02 & $-0.006$ & 0.006 & 0.99 & 0.02 & 0.969 & 0.003 & 7.1 & 6.4 & 7.8 \\
79578 & 145825 & 5820 & 5 & 4.47 & 0.01 & $0.057$ & 0.005 & 1.04 & 0.01 & 1.052 & 0.007 & 2.3 & 1.4 & 2.8 \\
79672 & 146233 & 5814 & 3 & 4.45 & 0.01 & $0.056$ & 0.003 & 1.02 & 0.01 & 1.045 & 0.005 & 3.0 & 2.4 & 3.3 \\
79715 & 145927 & 5803 & 6 & 4.38 & 0.02 & $-0.041$ & 0.005 & 1.09 & 0.01 & 0.998 & 0.004 & 6.4 & 5.9 & 6.8 \\
81746 & 150248 & 5715 & 5 & 4.40 & 0.02 & $-0.086$ & 0.004 & 0.99 & 0.01 & 0.960 & 0.002 & 7.6 & 6.8 & 8.0 \\
83276 & 153631 & 5885 & 8 & 4.22 & 0.02 & $-0.089$ & 0.006 & 1.23 & 0.01 & 1.038 & 0.008 & 7.4 & 7.1 & 7.7 \\
85042 & 157347 & 5694 & 5 & 4.41 & 0.02 & $0.015$ & 0.004 & 1.00 & 0.01 & 0.975 & 0.005 & 6.7 & 5.9 & 7.2 \\
87769 & 163441 & 5807 & 6 & 4.40 & 0.02 & $0.041$ & 0.006 & 1.05 & 0.01 & 1.024 & 0.005 & 5.2 & 4.3 & 5.7 \\
89650 & 167060 & 5841 & 5 & 4.44 & 0.02 & $-0.037$ & 0.005 & 1.08 & 0.01 & 1.026 & 0.006 & 3.8 & 3.0 & 4.4 \\
95962 & 183658 & 5806 & 5 & 4.44 & 0.02 & $0.023$ & 0.005 & 1.04 & 0.01 & 1.024 & 0.007 & 4.2 & 3.1 & 4.7 \\
96160 & 183579 & 5781 & 8 & 4.50 & 0.02 & $-0.053$ & 0.007 & 0.96 & 0.02 & 1.009 & 0.007 & 2.4 & 1.3 & 3.0 \\
101905 & 196390 & 5890 & 6 & 4.47 & 0.02 & $0.057$ & 0.006 & 1.07 & 0.02 & 1.072 & 0.008 & 1.1 & 0.8 & 2.3 \\
102040 & 197076A & 5838 & 6 & 4.48 & 0.02 & $-0.093$ & 0.006 & 1.05 & 0.02 & 1.018 & 0.008 & 2.9 & 1.7 & 3.5 \\
102152 & 197027 & 5718 & 5 & 4.40 & 0.02 & $-0.020$ & 0.005 & 0.95 & 0.01 & 0.972 & 0.004 & 6.7 & 6.0 & 7.4 \\
103983 & 200565 & 5752 & 10 & 4.51 & 0.02 & $-0.048$ & 0.008 & 0.96 & 0.02 & 1.000 & 0.008 & 1.9 & 1.0 & 2.8 \\
104045 & 200633 & 5831 & 6 & 4.47 & 0.02 & $0.045$ & 0.005 & 1.00 & 0.01 & 1.045 & 0.007 & 2.5 & 1.4 & 3.1 \\
105184 & 202628 & 5833 & 11 & 4.54 & 0.02 & $-0.002$ & 0.009 & 0.99 & 0.02 & 1.045 & 0.008 & 0.3 & 0.2 & 0.7 \\
108158 & 207700 & 5687 & 7 & 4.34 & 0.02 & $0.050$ & 0.007 & 0.97 & 0.02 & 0.980 & 0.003 & 8.3 & 7.8 & 8.7 \\
108468 & 208704 & 5829 & 7 & 4.33 & 0.02 & $-0.111$ & 0.006 & 1.16 & 0.01 & 0.990 & 0.003 & 7.5 & 7.1 & 7.9 \\
108996 & 209562 & 5847 & 17 & 4.53 & 0.03 & $0.064$ & 0.013 & 1.11 & 0.03 & 1.060 & 0.009 & 1.0 & 0.0 & 1.7 \\
109110 & 209779 & 5787 & 17 & 4.50 & 0.04 & $0.035$ & 0.014 & 1.06 & 0.03 & 1.026 & 0.011 & 1.8 & 1.0 & 3.4 \\
109821 & 210918 & 5746 & 7 & 4.31 & 0.02 & $-0.115$ & 0.005 & 1.06 & 0.01 & 0.960 & 0.003 & 9.2 & 8.8 & 9.6 \\
114615 & 219057 & 5816 & 9 & 4.52 & 0.02 & $-0.077$ & 0.008 & 1.04 & 0.02 & 1.022 & 0.008 & 0.5 & 0.2 & 1.6 \\
115577 & 220507 & 5699 & 9 & 4.25 & 0.03 & $0.036$ & 0.008 & 1.12 & 0.02 & 0.989 & 0.006 & 9.5 & 9.1 & 9.8 \\
116906 & 222582 & 5792 & 6 & 4.37 & 0.02 & $0.010$ & 0.005 & 1.05 & 0.01 & 1.003 & 0.005 & 6.5 & 6.0 & 6.8 \\
117367 & 223238 & 5871 & 8 & 4.32 & 0.02 & $0.044$ & 0.007 & 1.15 & 0.02 & 1.050 & 0.007 & 5.9 & 5.5 & 6.3 \\
118115 & 224383 & 5808 & 7 & 4.28 & 0.02 & $-0.017$ & 0.006 & 1.12 & 0.01 & 1.013 & 0.005 & 7.7 & 7.3 & 8.0 \\

\label{t:pars}
\end{longtable}
\end{landscape}
}

\end{document}